\pdfoutput=1
\documentclass[12pt,preprint]{aastex}

\slugcomment{ApJ, to be submitted, \today}

\usepackage{lineno}
\usepackage{rotating}
\usepackage{color}
\usepackage{graphics,graphicx}
\usepackage{pdfpages}
\usepackage{graphicx}
\usepackage{placeins}

\def\ltsima{$\; \buildrel < \over \sim \;$}
\def\simlt{\lower.5ex\hbox{\ltsima}} 
\def\gtsima{$\; \buildrel > \over \sim \;$}
\def\simgt{\lower.5ex\hbox{\gtsima}} 
\def\arcsec{\hbox{$^{\prime\prime}$}}
\def\deg{\hbox{$^\circ$}}

\def\Chandra{\textit{Chandra}}

\def\gb{GB~1428+4217}

\usepackage{amssymb,amsmath}
\newcommand{\Gam}{\hbox{\sc Gamma}}
\newcommand{\ind}{\buildrel{\rm indep}\over\sim}
\newcommand{\lira}{_{\rm LIRA}}
\newcommand{\yobs}{y_{\rm obs}}
\newcommand{\cur}{^{(t)}}
\newcommand{\ratxr}{\rho_{xr}\,}

\newcommand{\by}{\mbox{\boldmath{$y$}}}
\newcommand{\bLam}{\mbox{\boldmath{$\Lambda$}}}

\shorttitle{High-redshift Quasar X-ray Jets} 
\shortauthors{McKeough et al.}

\begin{document}

\title{Detecting Relativistic X-ray Jets in High-Redshift Quasars}

\author{
Kathryn McKeough\altaffilmark{1},
Aneta Siemiginowska\altaffilmark{2},
C.~C.~Cheung\altaffilmark{3}, \\
{\L}ukasz~Stawarz\altaffilmark{4}, 
Vinay L.\ Kashyap\altaffilmark{2},\\
Nathan Stein\altaffilmark{5}, 
Vasileios Stampoulis\altaffilmark{6},
David ~A. van Dyk\altaffilmark{6}, \\
J.~F.~C.~Wardle\altaffilmark{7},
N.~P.~Lee\altaffilmark{2},
D.~E.~Harris\altaffilmark{2}\altaffilmark{*},
D.~A.~Schwartz\altaffilmark{2},
Davide Donato\altaffilmark{8},\\
Laura Maraschi\altaffilmark{9} 
and
Fabrizio Tavecchio\altaffilmark{9}.
\\
}

\email{kathrynmckeough@g.harvard.edu}

\altaffiltext{1}{Department of Statistics, Harvard University, Cambridge, MA 02138}

\altaffiltext{2}{Harvard-Smithsonian Center for Astrophysics, 60 Garden St., 
Cambridge, MA 02138, USA}

\altaffiltext{3}{Space Science Division, Naval Research Laboratory, Washington, 
DC 20375-5352, USA}

\altaffiltext{4}{Astronomical Observatory, Jagiellonian University, ul. Orla 
171, 30-244, Krak\'ow, Poland}

\altaffiltext{5}{Department of Statistics, The Wharton School,
  University of Pennsylvania, 400 Jon M. Huntsman Hall, 3730 Walnut
  Street, Philadelphia, PA 19104-6340, USA}

\altaffiltext{6}{Statistics Section, Imperial College London, Huxley
  Building, South Kensington Campus, London SW7, UK}
 




\altaffiltext{7}{Department of Physics, MS~057, Brandeis University, Waltham, 
MA 02454, USA}



 \altaffiltext{8}{CRESST and Astroparticle Physics Laboratory NASA/GSFC, Greenbelt, MD 20771, USA}

\altaffiltext{9}{INAF Osservatorio Astronomico di Brera, via Brera 28, 20124,
Milano, Italy}

\altaffiltext{*}{Deceased formally from Harvard-Smithsonian Center for Astrophysics}

\begin{abstract} 
We analyze \Chandra\ X-ray images of a sample of 11 
quasars that are known to contain kiloparsec scale radio jets.
The sample consists of five high-redshift ($z\geq3.6$) flat-spectrum radio quasars,
and six intermediate redshift ($2.1<z<2.9$) quasars.  The dataset includes
four sources with integrated steep radio spectra and seven with flat
radio spectra.  A total of 25 radio jet features are present in this
sample.  We apply a Bayesian multi-scale image reconstruction method
to detect and measure the X-ray emission from the jets.  We compute deviations from a
baseline model that does not include the jet, and compare observed
X-ray images with those computed with simulated images where no jet
features exist. This allows us to compute $p$-value upper bounds on the
significance that an X-ray jet is detected in a pre-determined region of interest.  We
detected 12 of the features unambiguously, and an additional 6
marginally.  We also find residual emission in the cores of 3 quasars and in the background of 1 quasar
that suggest the existence of unresolved X-ray jets.  The dependence
of the X-ray to radio luminosity ratio on redshift is a potential
diagnostic of the emission mechanism, since the inverse Compton
scattering of cosmic microwave background photons (IC/CMB) is thought
to be redshift dependent, whereas in synchrotron models no clear
redshift dependence is expected.  We find that the high-redshift jets
have X-ray to radio flux ratios that are marginally inconsistent with those
from lower redshifts, suggesting that either the X-ray emissions is due to
the IC/CMB rather than the synchrotron process, or that high redshift jets
are qualitatively different.

\end{abstract}
 

\section{Introduction} 

Jets in active galactic nuclei transfer the energy generated by the central supermassive black hole (SMBH) 
to large( $>100$ kpc) distances.
The impact of jets on the environment contributes to the formation and evolution of structures in the
early Universe \citep{croton2006}. The innermost jets (parsec-scales or smaller) of radio-loud quasars are highly relativistic and their observed radiation can be Doppler amplified when observed at small angles to the line of sight. 
These jets can be bright at high-energies and thus can provide interesting observational probes of the state of the SMBH activity \citep{begelman84}, but they remain spatially unresolved in X-rays and $\gamma$-rays.

Large scale X-ray jets span distances out to hundreds of kiloparsecs away from the SMBH
and encode the history of SMBH activity during the jet's lifetime (a few Myrs). 
Their X-ray emission can be resolved with the \Chandra\ X-ray Observatory.
The number of such X-ray jets has significantly increased since the launch of \Chandra\ in 1999, but it is still relatively small in comparison to the number of known quasars.
There are about 100 large scale X-ray jets detected to date and only a few of them have good quality X-ray morphology data \citep{mas11}.
Though a direct connection between SMBH activity and the existence of kpc-scale jets is ambiguous, and the X-ray emission mechanism is not well understood, high-redshift ($z> 3$) jets could potentially establish the dominant energy environment in the early Universe.
Such jets probe the physics of the earliest (first $\sim$2~Gyr of the Universe in the quasars studied) actively accreting SMBH systems and are also interesting for other reasons.
For instance, the ambient medium in these high-redshift galaxies is probably higher than that of lower redshift galaxies \citep[e.g.,][]{dey06} and this may manifest itself in jets with different morphologies, with increased energy dissipation, or with the jets being slower in general than their lower-redshift counterparts.

The X-ray radiation could be attributed to either synchrotron emission
by highly relativistic electrons (Lorentz factors of $\gamma \sim 10^7-10^8$) in relatively strong magnetic fields, or
inverse Compton scattering of the cosmic microwave background (IC/CMB)
photons off the low energy ($\gamma \sim 10^3$) large-scale jet particles \cite[for a review see][]{har06}.  In
the simplest scenario, such models have diverging predictions at high
redshift.  Specifically, we expect a strong redshift dependence in the
X-ray--to--radio- energy flux ratio, $\ratxr = \frac{F_{\rm x}}{\nu_{\rm r} f_{\rm r}}$, where the radio fluxes are given at the observed frequency, and the X-ray fluxes are modeled over the energy range $0.5-7$~keV \citep{mas11}.
Typically, $\ratxr \propto U_{\rm CMB} \propto~(1+z)^{4}$ for IC/CMB, whereas in
synchrotron models, we do not expect a strong dependence,\footnote{The
energies of the synchrotron emitting electrons are different in the
observed radio and X-ray spectra, $\gamma_{\rm r} \sim 10^3$ vs. $\gamma_{\rm x}
\sim 10^7$ and these electrons may originate in the same population or
two different populations. Therefore, there could be some weak
redshift dependence in the synchrotron model.} $\ratxr \propto (1+z)^{0}$. Below we compare the predictions of these
two models for the highest-redshift relativistic jets.

Most \Chandra\ studies of quasar jets have so far targeted known arcsecond-scale radio jets \citep[e.g.,][]{sam04,mar11}, as most known examples are at $z$ $\stackrel{<}{{}_\sim}$2 \citep{bri84,liu02}.
At the time our program began, there were two high-$z$ quasars with kpc-scale X-ray jets: GB~1508$+$5714 at $z=4.3$ \citep{sie03,yua03,che04} and 1745$+$624 at $z=3.9$ \citep{che06}.
They were observed to have large $\ratxr$ values consistent with the IC/CMB model \citep{sch02,che04}, although the small number of high-$z$ detections precluded any definitive statements \citep{kat05,che06}.

We have therefore obtained \Chandra\ X-ray observations of an additional four high-redshift ($z>3.6$; GB\,1508$+$5714 was previously analyzed) and six intermediate-redshift ($2 \leq z \leq 3$) quasars with known radio jets.
The highest redshift X-ray and radio jet discovered in the sample studied (at $z=4.72$, in GB~1428+4214 (1428+422)) was presented and discussed in detail by \cite{che12}.
New and archival arcsecond-resolution NRAO\footnote{
The National Radio Astronomy Observatory is operated by Associated Universities, Inc.\ under a cooperative agreement with the National Science Foundation.}
imaging observations of these quasars are also presented.

The small number of X-ray photon counts observed from jets relative to their corresponding quasar cores means that detecting X-ray jets is inherently challenging.
Statistically, we must test the hypothesis that a baseline model of the quasar core and a flat background, without a jet, is insufficient to explain the observed data.
We do this test using a multi-scale Bayesian method known as Low Count Image Reconstruction and Analysis\footnote{
LIRA is implemented as a package for the R statistical programming language (r-project.org) that is available for downloading and use at {\tt github.com/astrostat/LIRA}.
} \citep[LIRA;][]{esch04, con07}.
The algorithm models the residual as a multi-scale component, and generates a series of images that capture the emission that may be present in excess of the baseline model.
We can then compute a $p$-value\footnote{
Formally, a $p$-value is the probability that the baseline null hypothesis can generate
 a value  for the test statistic 
as large as that which is observed.
In this case, it defines the likelihood that a given
intensity can be obtained under the assumption that the baseline model is the truth.
That is, when the $p$-value is small, the chances that the feature under consideration can be attributed to a fluctuation is small.
This allows us to {\sl reject} the null hypothesis when this probability falls below a pre-defined threshold.
Note, however, that it should never be interpreted as a measure of the probability that the alternate hypothesis is true, nor, if the null cannot be rejected, as a measure of the probability that the null hypothesis is true \citep{wl16}.
}
by generating a series of Monte Carlo simulations of images under the baseline model and fitting each of these simulated images using LIRA.
\cite{stei15} (hereafter Paper~I) show how an upper-bound on the $p$-value can be computed with a small number of MCMC replicates.
We are interested in detecting whether X-ray jets exist in regions where jets were previously observed in the radio band.
In this paper, we will not consider X-ray detections without a corresponding radio emission (such a detection of a jet that was recently reported by \cite{simionescu2016}) when matching our results to the IC/CMB or synchrotron emission model.
We run LIRA to detect jets in pre-defined regions of an X-ray image.
Using the jets detected in X-rays we are able to observe how $\ratxr$ is dependent on redshift and whether it matches the predictions of the IC/CMB or synchrotron emission model.

Section 2 describes the sample selection and initial processing of the X-ray and corresponding radio observations.
Section 3 outlines how LIRA is used to find evidence that a jet exists in a region where one  is observed in radio imaging.
Section 4 elaborates on the results of the image analysis methods when applied to the new X-ray observations.
Section 5 gives a final description of our results in context and we summarize our results in Section 6.

\section{Sample Selection and Observations}

Compilations of known radio jets \citep{bri84,liu02} show very few examples with kpc-scale extensions at $z>3$ .
To increase the number of known radio jets, we carried out a VLA survey of the highest redshift (z$>$3.4) quasars and searched for extended radio emission \citep{che05,che08}.
The sample sources are the brightest (\simgt 100 mJy at 1.4 and/or 5\,GHz to facilitate ease of radio mapping), flat-spectrum radio sources catalogued by NED.
The flat spectra is a good proxy for high beaming and our high redshift sample is 
representative of a general radio-loud quasars population.

Four flat-spectrum radio jets from the $z>3.4$ radio sample were observed in a \Chandra\ AO8 program (PI: Cheung) from 2007 Jan - June (ObsIDs: 7871-7874):
1239$+$376 ($z=3.82$),
1754$+$676 ($z=3.60$),
1418$-$064 ($z=3.69$),
1428$+$422 ($z=4.72$),
in addition to one quasar at $z=2.10$ (0833$+$585 ; ObsID 7870)
which had a known long radio jet \citep{mur93}.
The jets have minimum projected lengths of 2.5\arcsec\ and up to $\sim$15\arcsec\ for the $z=2.1$ case (Table~1), so their X-ray counterparts are easily separable from the bright nucleus in the \Chandra\ images.
The exposures of $\sim 3.8 - 11$ ks were tailored to the radio jet brightness and redshifts.

We also undertook a \Chandra\, survey of an intermediate redshift ($2 \leq z \leq 3$) sample of five sources in AO10 (PI: Sambruna) from  Jan - May 2009 (ObsIDs: 10307-10311) with exposures set to $\sim 20$ ks.
These radio jets were selected to have lengths greater than 2.5\arcsec\ from the list of \citet{liu02}.
In order to sample a range of jet orientations with respect to our line of sight, the selected targets have a range of radio core to extended flux ratios indicating both core-dominated and lobe-dominated radio sources.
A summary of the basic properties of the sources are recorded in Table~\ref{table-1}.

\subsection{Very Large Array Observations\label{sec:vla}}

The details of the VLA observations for the entire sample are shown in Table~\ref{table-2k}.
These radio jets form the largest sample of the radio jets at high redshift that were observed with {\it{Chandra}}.
We analyzed the archival data and selected the observations with the best astrometry for this project.
Portions of these radio data were presented in \citet{gob11} and \citet{gob14}.

The standard calibration was applied using AIPS \citep{bri94} with scans of primary calibrators, 3C~48 or 3C~286, used to set the flux density scales.
The data was then exported to the Caltech DIFMAP package \citep{she94} for self-calibration and imaging.

The radio jets in 1418$-$064, 1428$+$422, and 1754$+$676 were discovered in VLA 1.4 GHz A-configuration images from Oct - Dec 2004 (program AC755) and discussed in \citet{che05}.
We utilized the archival VLA snapshots of 1239$+$376 \citep{tay96} and 0833$+$585 \citep{mur93}.
The details of the 1428$+$422 observations are described in \citet{che12} but are included in Table~\ref{table-2k} for completeness.

The quasar 1418-064 has a hint of a radio extension in the VLA 1.4 GHz observation thus we obtained new A-array data at 5 GHz (July 27-28, 2007; program S8723) of this object which revealed more detail in the radio jet.

\subsection{\Chandra\ X-ray Observations.}

All X-ray observations were made with the \Chandra\ \citep{wei02, sch14} ACIS-S back-illuminated CCD.
The \Chandra\ observations are listed in Table~\ref{table-chandra}. 
We used the nominal aimpoint of the ACIS-S3 chip and a 1/8th subarray mode (0.4s frame time) in order to mitigate pileup of the nucleus.
Roll angles were selected to place a possible charge transfer streak from the X-ray nucleus away from the jet position angles.
We reprocessed the data by using {\tt chandra\_repro} script in CIAO \citep{fru06} and assigned the 
calibration available in CALDB v4.6.  We inspected the data for any
possible background flares and concluded that the data were not
affected by flares.  The background level in our short observations
was low and detections of the quasar emission and some jet features
are highly significant. We only use the X-ray events with
energies between 0.5-7\,keV in the spectral and image analysis
described below.

The X-ray nuclei are all clearly detected and the \Chandra\ positions (peaks) are within the 0.6\arcsec\ (90$\%$ pointing accuracy \citet{wei02}) of the radio positions (Table~\ref{table-1}).
We use the method described below for image analysis and detections of any features outside the core.


All spectral modeling was performed in Sherpa \citep{fre01} CIAO
version v4.6.  We extracted the spectra and created response files for
each observation using CIAO tools. 
We used the Nelder-Mead optimization algorithm and Cash likelihood
statistics appropriate for low counts data and fit the spectra in
Sherpa.  We assumed an absorbed power law model for each identified
feature with the absorption column fixed at the appropriate Galactic
value (Colden; \cite{stark1992}).  The photon index of the power law and
the normalization were fit.  For the cores, we also fit a model which
included an additional intrinsic absorption component at the quasar
redshift.

\section{X-ray Image Analysis Methods}

\subsection{Initial Analysis}

For each \Chandra\ observation we generate an image centered on the quasar rebinned to a pixel size of $0.246$\arcsec (half of the native ACIS pixel size.)
For sources with a radio jet located at $< 7$\arcsec\, from the quasar core we use $64\times 64$ pixel ($15.7$\arcsec$\times 15.7$\arcsec) X-ray images.
For the three sources with larger radio structures we use $128 \times 128$ pixels ($31.5$\arcsec$ \times 31.5$\arcsec) images.

We employ SAOtrace\footnote{CXC Optics: http://cxc.harvard.edu/cal/Hrma/SAOTrace.html} to generate \Chandra\ point-spread functions (PSFs) for each observation, constructing the PSF with parameters appropriate to the quasar location.  
Since X-ray PSFs are energy dependent, the raytraced PSFs include spectral models derived from fitting the spectrum of each quasar.
We use the same pixel size as in the X-ray image when binning the PSF.

We select regions of interest (ROI) for the quasar core and their jets using existing radio images, extracting spectra and response files using {\tt specextract}.
The radio images appear in panel (a) of Figure~\ref{fig:10307} and of Figures~\ref{fig:10308}--\ref{fig:2241} in Appendix B and the ROIs are overlaid on \Chandra\ ACIS-S images in panel (b) of each figure.
The circular quasar regions are marked `Q' and are centered on the radio core with the intent to encapsulate extended regions around the core, and the elliptical ROIs are associated with jet features and are numbered.
The complementary region is the remainder of the image that excludes both the quasar and jet regions and are marked 'C'.
The number of ROIs vary from one quasar to another.
Spectra are extracted from these regions and fit with an absorbed power-law model.
The results are in Table~\ref{tbl:jets}.
\begin{figure*}[ht]
\centering
\begin{tabular}[b]{@{}p{0.3\textwidth}@{}}
\centering\small (a)
\centering\includegraphics[scale=0.205,trim= 120 0 150 32,clip]{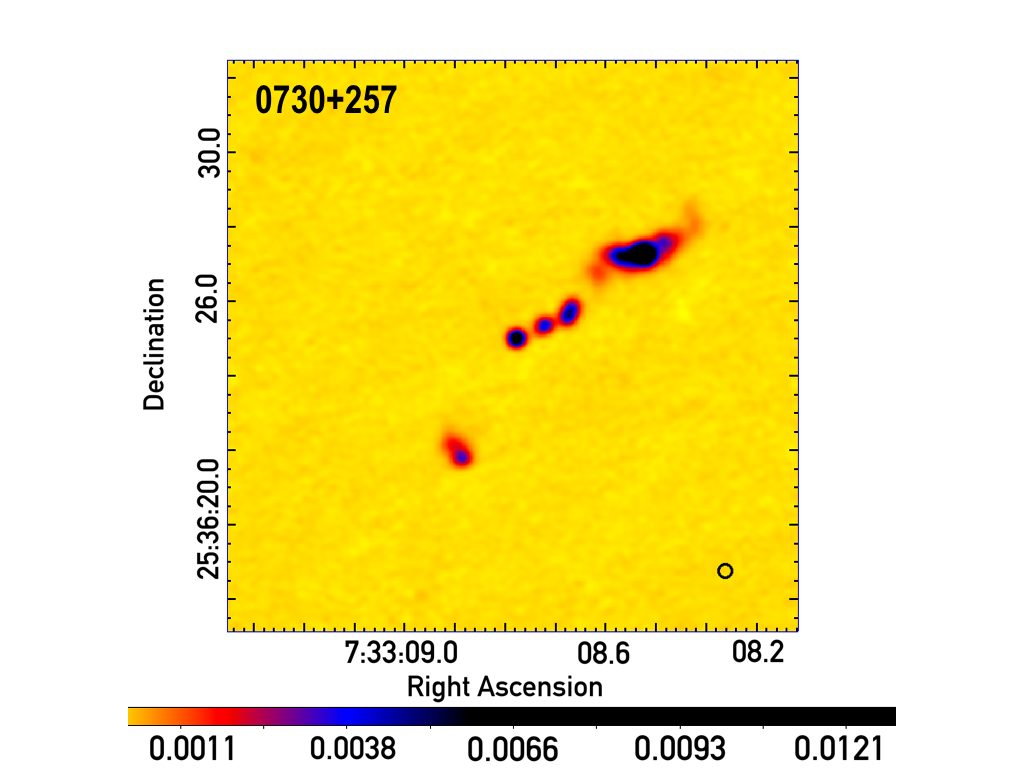}\\
\end{tabular}
\quad
\begin{tabular}[b]{@{}p{0.3\textwidth}@{}}
\centering\small (b)
\centering \includegraphics[scale=0.205,trim= 14 0 10 23,clip]{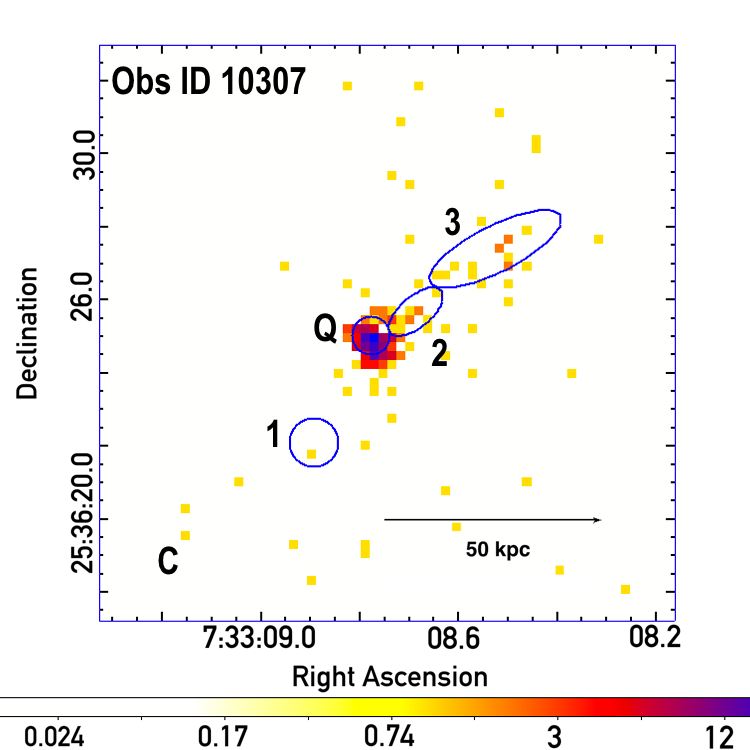}\\
\end{tabular}
\quad
\begin{tabular}[b]{@{}p{0.3\textwidth}@{}}
\centering\small (c)
\centering\includegraphics[scale=0.2,trim= 120 0 170 20,clip]{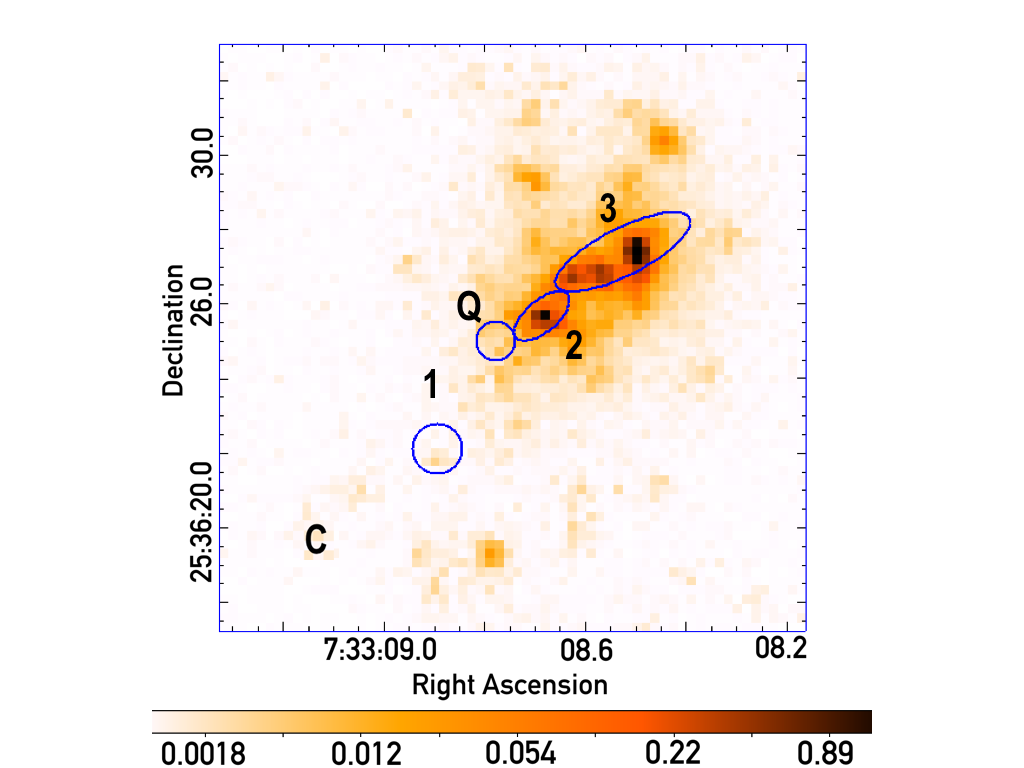}\\
\end{tabular}\\
\begin{tabular}[b]{@{}p{\textwidth}@{}}
\centering\small (d) \\
\centering\includegraphics[scale=0.5]{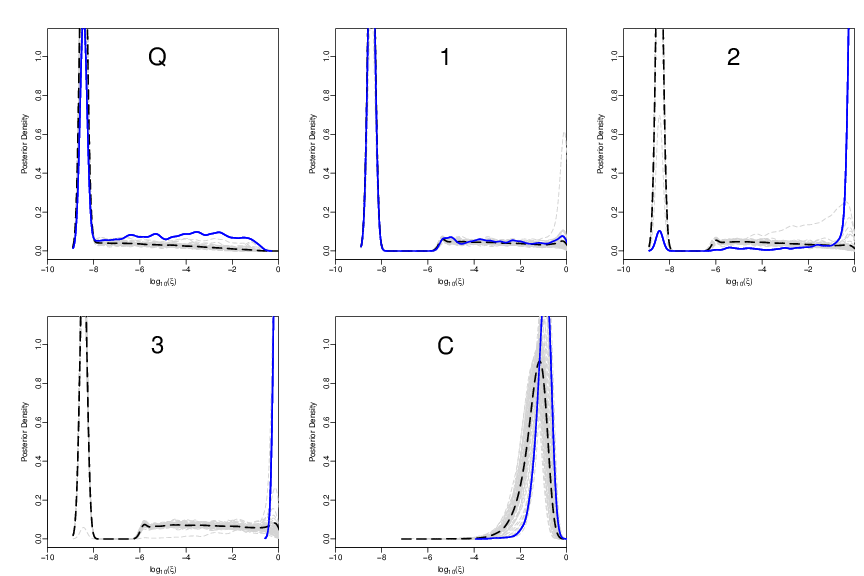}
\end{tabular}
\caption{\small Images, regions of interest (ROIs), and results for quasar 0730$+$257 (ObsID 10307) -- (a) The VLA radio image with colors indicating the intensity in janksy per beam. The beam size is shown as an ellipse in the bottom right corner of the panel.-- (b) The observed \Chandra\ ACIS-S image with colors indicating photon counts. -- (c) The fitted added structure with colors indicating the residual intensity calculated from the posterior mean of LIRA ($\tau_1\Lambda_1$). -- (d.Q)-(d.C) The posterior distributions of $\xi$ in each ROI for the data (blue solid curve), the $50$ simulated replicate images under the null model, which includes the quasar and a background but not a jet (grey solid curves), and the average across them (black dashed curve). The ROIs are shown superimposed on panels (b) and (c). }
\label{fig:10307}
\end{figure*}
\FloatBarrier
 
\subsection{Image Reconstruction using LIRA}

Detection of an X-ray jet is particularly challenging in part because it is a faint structure located near a much stronger quasar core.
The Bayesian multi-scale fitting method LIRA is well suited to this challenge.
An efficient feature detection method is described in Paper~I as an addendum to the multi-scale fitting.
Section~2 of Paper~I gives full details of the statistical models used by LIRA, which we briefly review here.
We consider square images of $n=2^d{\times}2^d$ pixels, where $d$ is an integer (for typical cases considered here, $d=6$ or $7$).
We denote the counts in each pixel by $\by = (y_i, i=1 \ldots n)$, with the total number of photon counts, $N = \sum_i y_i$.
The image is then modeled as the sum of a known {\sl baseline component}, here representing the quasar core and background, and an {\sl added multi-scale component}, here representing the jet.
The counts observed in detector pixel $i$ are modeled as
\begin{equation}\label{eq:detected}
y_i \ind  \textrm{Poisson}\left(\sum_{j=1}^n P_{ji}A_j\left(\tau_0\Lambda_{0j} + \tau_1\Lambda_{1j}\right)\right) \,,
\end{equation}
where $(P_{ji}, i=1 \dots n, j=1 \dots n)$ is the PSF, representing the probability that a photon arriving from the direction represented by pixel $j$ is recorded in pixel $i$,
$(A_j, j=1 \dots n)$ is the exposure map, representing the efficiency with which incoming photons at pixel $j$ will be recorded,
$\tau_1$ is the total intensity of the added component,
and $\tau_0$ is the total intensity of the baseline component.
The quantities $\bLam_k = (\Lambda_{k1},\ldots,\Lambda_{kn})$ (for $k=0,1$) are the proportions of $\tau_k$ in the individual pixels.
Note that this represents a general model that is applied to the full image.
For ROIs smaller than the full image (see below), we first apply the model in Equation~\ref{eq:detected} to the full image, then recompute $\tau_k = \sum_{j{\in}{\rm ROI}} \mu_{kj}$, where $\mu_{kj}=(\tau_k\Lambda_{kj})$ are the inferred intensities in each pixel.

Paper~I describes how LIRA uses a flexible multi-scale model in order to accommodate added structure, such as a jet, in the quasar image.
LIRA combines the likelihood given in Equation~\ref{eq:detected} with prior distributions to formulate a posterior distribution for $\tau_0$, $\tau_1$, and $\Lambda_1$ and uses MCMC to provide a Monte Carlo sample from this posterior distribution.

\subsection{Using LIRA to Quantify Evidence for Extended Jet Emission}

LIRA uses a binned \Chandra\ image, a ray traced \Chandra\ PSF, and a composite Gaussian plus constant baseline model representing the quasar core and background respectively.
The MCMC sampler within LIRA returns a sequence of simulated images of the residual multi-scale component, including any non-Poisson deviations from the baseline model, which in the cases studied, would indicate a jet emanating from the X-ray quasar..
If the \Chandra\ image is consistent with the baseline model (quasar core + background), the fitted image of the multi-scale component contains only random fluctuations.
On the other hand, any X-ray emission not described by the baseline component appears as a systematic deviation from this noise.
Substantial deviations can be taken as evidence for structures beyond the baseline component in the \Chandra\ image.

Section~3
of Paper~I describes a formal statistical hypothesis test for the presence of structure (e.g., a quasar jet) in the added multi-scale component and how we can compute an upper bound on the $p$-value for this test.
This involves Monte Carlo evaluation of the behavior of the fitted multi-scale component (which describes the deviation from the baseline image) when there is no actual structure in the X-ray image beyond the baseline model (i.e., when there is no jet).
Quantitatively, this involves using a test statistic to efficiently compute an upper bound on the $p$-value.
Here we outline this method; further details can be found in Paper~I.

To start, for each observed image we simulate $50$ replicate images under a baseline-only model using Sherpa's {\tt fake} function.
The baseline model is composed of a 2D Gaussian for the quasar core on top of constant background and set to the best-fit values determined by fitting a 2D model, {\tt psf*gauss2d+const2d}, to the observed X-ray image.
For each quasar, we run LIRA \footnote{We only use LIRA iterations that are obtained after an initial burn in of the MCMC sampler.
In all cases, we obtain 2000 iterations for each run} and discard the first 1000
on the \Chandra\ images as well as each of its 50 replicate baseline-only simulated images.
The image of the deviation from the baseline when fit to the simulated images contain only random fluctuations, while those fit to the \Chandra\ images contain estimates of any structure beyond the baseline model.
A jet is considered detected if the signal in the region of interest is stronger than is likely to occur through random fluctuations under the baseline-only model.
Thus, we must compare the fitted deviations from the baseline of the \Chandra\ observation with the corresponding $50$ fitted deviations of the baseline-only simulations.

This is done quantitatively through a novel test statistic that compares the expected photon counts due to the baseline and added components, i.e., it compares $\tau_0$ with $\tau_1$ in Equation 1.
The test statistic is unusual in that it is a posterior probability, namely, 
\begin{equation}
T_c(\by) = \Pr( \xi \geq c \mid \by),
\end{equation}
given a threshold $c$ where
\begin{equation}
\xi =\tau_1/(\tau_1 + \tau_0)
\end{equation}
is the proportion of expected counts due to the added structure.
If there is no structure in the added component (e.g., no quasar jet), $\xi$ and thus $T_c(\by)$ are close to zero; as the intensity of the added component grows, $\xi$ and $T_c(\by)$ tend toward one.
In this regard, $T_c(\by)$ is a useful statistic for testing for added structure, i.e., for the quasar jet.


\section{Results}

\subsection{Quasars}

The quasar nuclei in our sample were observed by \Chandra\ with photon counts ranging from 64 to 5500.
We obtained a relatively good spectrum for each quasar core, and fit a power-law with a fixed Galactic absorption model to each spectrum using Sherpa \citep{fre01}.
We also considered an additional intrinsic
absorption component ({\tt
xsphabs(Gal)*xszphabs($z_{qso}$)*powlaw1d}).  The results of our
spectral fits are shown in Table~\ref{tbl:quasar2}.  An intrinsic
absorption is required in four quasars (0730+257, 1218+112, 1834+612, and 1428+422), with absorption columns of
${\approx}10^{22}$~cm$^{-2}$. For the remaining seven quasars we obtained upper limits. The photon indices for the power law
model range between $\Gamma_X=1.35$ and $\Gamma_X=2.26$
with an average of $\Gamma_X = 1.7\pm0.2$, consistent with estimates for large
samples of radio-quiet and radio-loud quasars \citep{kelly2007, young2009, lanzuisi2013}.  All the quasars
are X-ray luminous, with the luminosities in $2-10$\,keV energy band
exceeding $2.5\times 10^{45}$~erg~s$^{-1}$.
The most luminous quasar in our sample (1428$+$422) has a $2-10$\,keV luminosity of $4.0\pm 0.2 \times 10^{47}$~erg~s$^{-1}$. 

\subsection{Jets}

We applied the method presented in Section 3 to our sample of 11 high-redshift quasars.
The posterior mean of $\tau_1 \Lambda_1$ is plotted in panel (c) of Figure~\ref{fig:10307} and of Figures \ref{fig:10308} - \ref{fig:2241} in Appendix B.
These represent the fitted multi-scale structure added to the baseline, i.e., the jets.
In order to assess the significance of the X-ray emission in each ROI and claim a detection, we calculate an upper bound on the $p$-value (with an upper tail probability of $\gamma = 0.005$, see section 3.4 of Paper~I).
This requires us to post-process the LIRA MCMC posterior sample to obtain a posterior sample of $\xi$ for each ROI.
Panel (d) of Figure~\ref{fig:10307} and of Figures~\ref{fig:10308}--\ref{fig:2241}  in Appendix B display the smoothed distributions of $\xi$ for each ROI for the data (solid blue line), and the averaged distributions of $\xi$ over the same ROI for the null simulations (dashed black).
The distributions of $\xi$ for each of the $50$ simulated null images are shown as grey lines.
We expect the observed distribution to depart from the distributions based on the null simulations if there is significantly more X-ray emission than expected under the baseline model.

We set a threshold of $p=0.01$ to determine the existence of a significant feature.
That is, if the upper bound on the $p$-value for the distribution of $\xi$ in an ROI is ${\leq}0.01$, we take that to be sufficient evidence for X-ray emission associated with a jet feature in the ROI.
Table~\ref{tbl:sigx} lists the upper bounds on the $p$-value for all the ROIs, showing that 16 features (12 jet features, 3 cores and 1 complementary region) have a $p$-values less than 0.01.
There are an additional 7 ROIs with an upper bound on the $p$-value between 0.01 and 0.02 which we consider marginal detections.
Thus, in total we list 23 significant X-ray features associated with high redshift jets including the corresponding cores and complementary regions.
In all these cases the distributions of $\xi$ based on the \Chandra\ data are skewed to the right compared to the distributions based on the null simulations.

There are three ROIs associated with the quasar cores (region Q for sources 1508$+$5714, 1428$+$422 and 1834$+$612) for which the upper bound on the $p$-value is lower than our pre-defined $p$-value threshold of $\alpha=0.01$.
This indicates the detection of the X-ray emission in excess of a point source assumed for the spatial model of the quasar core emission.
In general, all the images representing the deviation from the baseline model (panel (c) of Figure~\ref{fig:10307} and of Figures~\ref{fig:10308}--\ref{fig:2241} in Appendix B ) show indications of excess counts in the core ROIs in all the sources; we attribute these to an imperfect knowledge of the true telescope PSF, with the apparent deviations reflecting the uncertainties in the adopted PSF.
However, only in three quasars are they significant, signaling a departure from the baseline model. Note that this also shows that our threshold for detection is stringent enough to avoid detecting spurious features that may arise due to uncertainties in the PSF.

Note that the in the case of 1508$+$5714, a significant secondary structure in the complementary region `C' is seen, implying that not all the X-ray features were captured by the radio-based regions.
This object has a deep \Chandra\ observation of about 90~ksec, the longest one in our sample, and it is not unusual to find such structure in long observations.

Our X-ray jets detection rate -- 9 sources with detected jets out of 11 with known jets --
is similar to the one reported for the lower redshift sources ($\sim 70\%$ detection rates for exposures of $\sim$5--10 ksec; \cite{sam04,mar11}).

\section{Discussion}

\subsection{X-ray Morphology}

The posterior mean images $\tau_1 \Lambda_1$ representing deviations from the baseline model of the \Chandra\ data (panel (c) of Figure~\ref{fig:10307} and of Figures~\ref{fig:10308}--\ref{fig:2241} in Appendix B provide a view of the X-ray morphology with the quasar core removed.
However, not all the structures seen in these images are significant, since many features could be attributed to statistical fluctuations.
We have developed a method (Paper~I; see also Section 2 above) to assess the significance of the emission for well-defined ROIs.
Note that these ROIs {\sl must be set prior to the analysis} and cannot be deduced from the LIRA output, since doing so would increase the false detection rate.
We adopt regions based on the locations of the radio jet features, but note that the jet X-ray emission may not always be spatially coincidental with the radio emission \citep{sch02}, nor even have a radio counterpart \citep{jor04, simionescu2016}.
Offsets between the radio and X-ray peaks in the jet features have been reported \cite[e.g.][]{sie2007, worrall2009}. 
For instance, in 1754$+$676 (ObsID 7872), the X-ray jet is not detected, but the image showing deviations from the baseline (see panel (c) of Figure~\ref{fig:7872}) suggests the existence of an emission feature between the quasar and the radio jet region.
A longer \Chandra\ observation is necessary to confirm this emission.
In another source 0805$+$046 (ObsID 10308), the image showing deviations from the baseline (panel (c) of Figure~\ref{fig:10308}) displays considerable emission outside the narrow radio jet, suggesting a more complex X-ray morphology.

We used the complementary regions to assess the possibility of unexpected X-ray emission present outside the pre-defined regions.
In all sources but 1508$+$5714, we do not find a strong indication that such emission is present, though the complementary regions cover a large area and thus statistical tests have relatively low power to detect smaller compact structures.
Future studies of the X-ray morphology in the vicinity of this source is required for understanding the origin of this emission.

The $p$-value upper bound test relies on the test regions being pre-defined.
This is done in order to avoid the loss of power in the test that arises when multiple hypotheses are tested.
We thus take the regions directly from the radio data and do not optimize the regions based on the X-ray data.
This could result in the size of location of the regions to be slightly misaligned, reducing the significance of detection.
For example, ROI 1 in source 1428$+$422 (ObsID 7874) is a marginal detection with an upper bound on the $p$-value of 0.010, but decreasing the region size from 91 to 77 pixels, an arguably better fit for this object in radio, results in an improvement in the $p$-value upper bound to 0.009, which crosses the threshold into a significant detection.
Areas of deviation from the baseline may be difficult to detect in a large region or a region may not encompass all of the relevant area of the image, and the better the region fits an area with deviation from the baseline, the lower the nominal $p$-value is.
But when large numbers of regions are tested, the $p$-value threshold must be reduced correspondingly in order to prevent false claims of detections due to fluctuations.
For instance, if the $\xi$ in 20 regions (say) of radii stepping from 75 to 95 pixels are tested, the appropriate threshold of $\alpha$ must be reduced by a factor of 20, to $\alpha=0.0005$, to maintain the same level of significance.
We emphasize that ROI selection must be consistent across the analysis, and {\sl must be defined before} applying the significance test.
We also note that we apply the test to a total of 47 ROIs, so we expect at most 1 false positive amongst the claimed significant and marginal detections at the significance threshold of $\alpha<0.02$.

As noted above, our current method of region selection depends on
radio data.  In the future, we plan to develop methods that are
independent of the radio (or other wavebands) selections and
autonomously generate regions that adaptively fit the deviations from
the baseline in the LIRA output.  Such a method is needed since the
X-ray emission does not always follow the radio closely.  Naturally,
any such method will trade-off ROI optimization for the statistical power of
the detection routine.

\subsection{Redshift Dependence in Large-Scale X-ray/Radio Emission}

The origin of X-ray jet emission is still under debate.
An early hint at the advantage of studying high-redshift jets came from the $z$=4.3 quasar 1508$+$5714 (Siemiginowska et al.\ 2003; Yuan et al.\ 2003). This quasar has higher X-ray to radio 
luminosity ratio ($\ratxr>$100) than any of its lower-$z$ counterparts (Cheung 2004).
This appears consistent with the $(1+z)^{4}$ amplification in the energy density of Cosmic Microwave Background (CMB):
\begin{equation}
\ratxr \propto u_{\rm CMB}/u_{\rm B} \propto (1+z)^{4}(\delta/B)^{2} \,, 
\end{equation}
as expected under the IC/CMB model (e.g., Schwartz 2002). 

\begin{figure*}[ht]
\centering
\includegraphics[width=0.9\columnwidth]{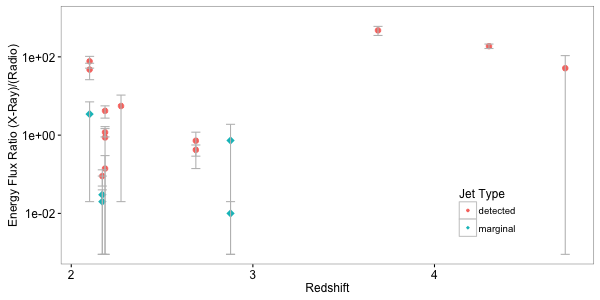}
\caption{
Ratio of X-ray to radio flux $\ratxr$ vs redshift for the detected and marginally detected regions of interest of the jets.
The circles are the energy flux ratios from jets detected in this study.
The diamonds are estimated energy ratios from the marginally detected jets.
The error bars form the $68\%$ interval from the LIRA iterations.
} 
\label{fig:fluxRatio}
\end{figure*}

\begin{figure*}[ht]
\centering
\includegraphics[width=0.7\columnwidth]{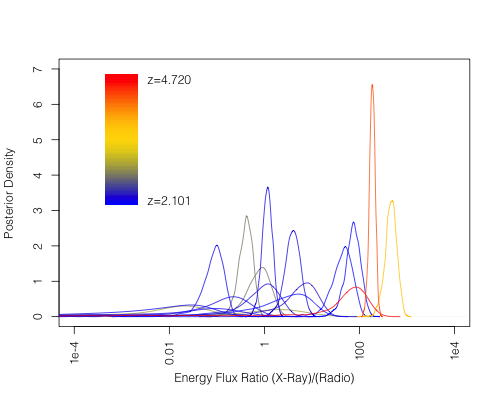}
\caption{
The posterior distribution of the ratio of X-ray to radio fluxes, $\ratxr$, for each detected and marginally detected jet.
The energy flux ratio is calculated at every iteration of LIRA.
The color corresponds to redshift.
}
\label{fig:density}
\end{figure*}

\begin{figure*}[ht]
\centering
\includegraphics[width=0.8\columnwidth]{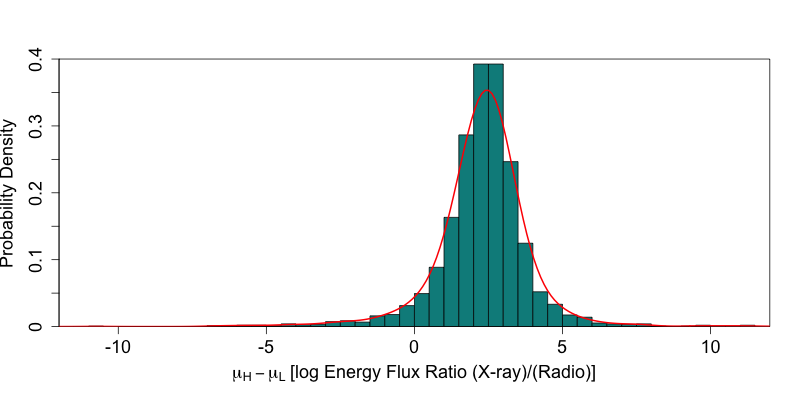}
\caption{
The difference in the mean X-ray to radio energy flux ratio between high- and medium-redshift quasars, $\mu_H - \mu_L$ at every LIRA iteration.
$\mu_L$ and $\mu_H$ are the average log flux ratio across the lower-redshift ($2<z<3$) and higher-redshift ($z>3$) detected and marginally detected redshift jets.
}
\label{fig:mustat}
\end{figure*}

Figure~\ref{fig:fluxRatio} compares the energy flux ratio $\ratxr = [F_{\rm x}~{(0.5-7\rm keV)}]/[\nu_{\rm r} f_{\rm r}]$ of the
detected and marginally detected jets across redshift from our
\Chandra\ sample.  We seek to establish whether or not the energy flux
ratio varies with redshift.  Figure~\ref{fig:density} shows the
posterior distribution from LIRA of the energy flux ratio for each
detected and marginally detected source (blue corresponds to low, and
red to high redshift).  It is visually apparent that there is a
difference in the distributions of the sources with higher redshifts
($z>3$) and those with lower redshifts.  In order to establish a
statistical measure of the significance of this difference, we split
the detected and marginally detected sources into two samples
consisting of the 18 at low redshift ($z<3$; sample $L$) and the 3 at
high redshift ($z>3$; sample $H$).  We then use a hierarchical Gaussian model to
examine whether low and high redshift quasars differ in terms of the
mean and variance of their energy flux ratio.  Appendix A describes a
procedure for evaluating the posterior probability that the difference
between the mean $\log_{10}$ energy flux ratio of the high and low
redshift jets ($\mu_H - \mu_L$) is greater than zero.
Figure~\ref{fig:mustat} shows the distribution of $\mu_H-\mu_L$
calculated from the posterior output.  We find an empirical
probability of $95\%$ that $\mu_H-\mu_L \geq 0$, which is at best
marginal evidence that the observed difference cannot be due to a
statistical fluctuation.  Though highly suggestive, because of the
small number of sources represented in this paper, and given the
disproportionate numbers of jets in the two samples, there is
insufficient evidence to conclude that the mean $\log_{10}$ energy
flux ratio differs between two groups of jets.  More observations at
$z>3$ are required in order to obtain more reliable results.

The difference between the jets in two redshift groups is interesting
because it can also indicate that radio-loud quasars at $z>3$ are
different from low redshift ones. \cite{volonteri2011} hypothesized
that the jets at $z>3$ are systematically slower in comparison to the
jets at $z<3$.  If true, this could affect the energy flux ratio in
the framework of the IC/CMB model, due to a strong dependence of this
ratio on the jet Doppler factor. In this case, assuming that the
comoving jet magnetic field is roughly the same at different
redshifts, the observed increase in the energy flux ratio should be
smaller than that expected from the $(1+z)^4$ scaling. However, the
jet magnetization may evolve with redshift \citep[e.g.][]{singal2013}
which complicates the redshift scaling even further. Our results
indicate that the high redshift jets are different, but more
observations are needed to study the origin of this difference.

\subsection{Quasars at High Redshift}

The X-ray emission of radio loud quasars observed with {\it Chandra}
is unresolved and contained within $< 1.5\arcsec$ circular regions.
This emission could be due to a mixture of at least three components:
a hot corona directly related to the accretion process, a parsec scale
jet, and an unresolved portion of the kpc-scale outflow emitting
X-rays via IC/CMB.  We measured a standard range of photon indices for
the assumed power law model for quasar core spectra and found them
consistent with either process.  However, we detected relatively low
values of photon indices in a few quasars, including two at the
highest redshift (see Table~\ref{tbl:quasar2}).  Lower values of the
photon index are predicted if the jet dominates the X-ray emission.
In this case the beamed jet would make the quasars to appear more
luminous.  We notice that such a trend is present in our small sample
and five sources with $\Gamma_X < 1.6$ are more luminous, with an
average 2-10~keV luminosity of $6.3 \pm 2.5 \times
(10^{46}$erg~s$^{-1})$, than the other six with $\Gamma_X>1.6$ and the
average luminosity of $7.9 \pm 2.5 \times (10^{45}$erg~s$^{-1})$.
Formal correlation tests show that $\Gamma_X$ and ${\log}L_X$ are
indeed correlated, with Pearson's correlation coefficient $\rho=0.61$
($p=0.047$), and Kendall's $\tau=0.53$ ($p=0.024$). Such relation has
been seen in analysis of large samples of radio-loud and radio-quiet
quasars \citep{bechtold1994,
young2009,lanzuisi2013}. \cite{bechtold1994} found a similar trend in
a sample of radio loud quasars and argued that it could be caused by
an increased absorption. On the other hand \cite{young2009} did not
find a significant correlation in a radio-loud subsample of quasars
observed with XMM-Newton. Future studies of large number of radio loud
quasars in X-ray and radio band are necessary for understanding the
presence and origin of this correlation.

In the analysis of the {\it Chandra} images we assumed that the quasar
emission is point-like.  However, we detect signatures of
non-point--like emission in the two highest redshift ($z>4$) quasars
(1428+422 and 1508+5714) and in the one at $z=2.3$ (1834+612) (as evidenced by the fact that the
core component Q is not fully accounted for in the baseline model, see
Table~\ref{tbl:jets} and
Figures~\ref{fig:10311},\ref{fig:7874},\ref{fig:2241}).  This is
unlikely to be due to uncertainties in the shape of the \Chandra\ PSF,
since the residual core emission is not present in the multi-scale
components for other sources.  We suggest that these are due to
non-negligible contributions from unresolved kpc-scale jets emitting
IC/CMB.  The effect of this residual component on our analysis is
conservative, i.e., the imperfect modeling of the core tends to
increase the strength of the baseline model and systematically dampen
the added multi-scale component.  
Thus, our results indicate that the unresolved X-ray cores of radio loud
high redshift quasars can contain significant contributions from
kpc-scale jet emission. This result is in agreement with studies based
on the optical-to-X-ray luminosity ratio \citep{saez2011,wu2013} and
along the expectations from the IC/CMB model.  Such a jet contribution
can potentially bias population studies of quasars and needs to be
taken into account in investigations of radio loud quasars at high
redshift.

\section{Summary}

We have analyzed \Chandra\ X-ray observations of a sample of eleven $z>2$ quasar radio jets including flat-spectrum and steep-spectrum radio sources.
The quasars were selected based on the known radio jets.
We use the Bayesian multi-scale image reconstruction method, LIRA, to obtain high-quality images outside the quasar core, and assess the significances of X-ray emission features coincident with jets.
We detect X-ray counterparts to radio jets in 9 quasars, including in the highest redshift X-ray and radio jet currently known (\gb\ at $z=4.72$, Cheung et al.(2012)).
In particular, 12 radio features are detected at high significance, and an additional 6 at marginal significance.

We find that the ratio of X-ray to radio energy flux may differ
between jets at high ($z>3$) and low ($z<3$) redshift, in accordance
with the IC/CMB mechanism for X-ray emission, or pointing to the
intrinsic differences between low and high redshift quasars.  However,
this difference is subject to a large uncertainty due to small sample
sizes, and more observations are needed in the high redshift regime to
confirm this trend.

\acknowledgments

We remember Dan Harris' passion for understanding jets:  ``After all these years and all these conferences on jets, we still don?t know what jets are made of or how they work" (Harris 2015).
The authors acknowledge Rita Sambruna's contribution as the PI of the Chandra X-ray Observatory program for a sub-sample of quasars analyzed in this work. A.S. thanks Francesco Massaro and Giulia Migliori for discussion. Doug Gobeille participated in the initial selection of the radio jets.
This work was supported in part by the National Science Foundation REU and Department of Defense ASSURE programs under NSF Grant no.\ 1262851 and by the Smithsonian Institution.
The methodological aspects of this project were conducted under the auspices of the CHASC International Astrostatstics Center.
CHASC is supported by NSF grants DMS 1208791, DMS 1209232, DMS 1513546  and 1513492.
DvD acknowledges support from a Wolfson Research Merit Award (WM110023) provided by the British Royal Society and from Marie-Curie Career Integration (FP7-PEOPLE-2012-CIG-321865) and Marie-Skodowska-Curie RISE (H2020-MSCA-RISE-2015-691164) Grants both provided by the European Commission.
This research was supported in part by NASA through contract NAS8-03060 (A.S., D.A.S., V.L.K.) to the \Chandra\ X-ray Center. It was also supported by the \Chandra\ Award number GO7-8114 to Brandeis University by the \Chandra\ X-Ray Observatory Center, which is operated by the Smithsonian Astrophysical Observatory for and on behalf of NASA under contract NAS8-03060.
L.S. was supported by Polish NSC grant DEC-2012/04/A/ST9/00083.

{\it Facilities:} \facility{CXO}, \facility{VLA}

{}


\newpage



\clearpage






\clearpage


\begin{table}
\caption[]{\label{table-1} Sample Summary of Basic Properties}
\begin{center}
\begin{tabular}{lcccccccc}
\hline \hline
Name & R.A.  & Decl.  & Ref. & $z$ & Ref. & $D_{\rm L}$ & Scale \\ &
(J2000.0) & (J2000.0) & & & & (Gpc) & (kpc/\arcsec) \\
\hline
0730$+$257   & 07 33 08.784 &  +25 36 25.06 & SDSS & 2.686 & O82 & 22.6 & 8.06 \\
0805$+$046   & 08 07 57.539 &  +04 32 34.53 & F04  & 2.877 & L70 & 24.6 & 7.92 \\
1311$-$270  & 13 13 47.401 & --27 16 49.13 & USNO & 2.186 & B02 & 17.6 & 8.39 \\
1318$+$113   & 13 21 18.838 &  +11 06 49.98 & SDSS & 2.171 & L72 & 17.4 & 8.39 \\
1834$+$612   & 18 35 19.675 &  +61 19 40.01 & B02  & 2.274 & H97 & 18.4 & 8.34 \\
0833$+$585   & 08 37 22.410 &  +58 25 01.85 & J95  & 2.101 & K80 & 16.7 & 8.43 \\
1239$+$376   & 12 42 09.812 &  +37 20 05.69 & F04  & 3.818 & V96 & 34.5 & 7.22 \\
1754$+$676   & 17 54 22.187 &  +67 37 36.36 & USNO & 3.60  & V99 & 32.2 & 7.38 \\
1418$-$064  & 14 21 07.756 & --06 43 56.36 & B02  & 3.689 & E01 & 33.2 & 7.31 \\
1428$+$422   & 14 30 23.742 &  +42 04 36.49 & F04  & 4.72  & H98 & 44.5 & 6.59 \\
1508$+$5714  & 15 10 02.922 &  +57 02 43.37 & M98  & 4.30  & H95 & 39.8 & 6.87 \\
\hline \hline                                          
\end{tabular}
\end{center}
Notes -- The quasar radio positions are taken from:
B02 = \citet{bea02}, 
F04 = \citet{fey04}, and 
J95 = \citet{joh95},
while the optical measurements are from SDSS DR6 \citep{ade08} and USNO B1.0 \citep{mon03}.\\
The redshifts ($z$) are from:
B02 = \citet{bak02},
E01 = \citet{ell01},
H97 = \citet{hen97},
H98 = \citet{hoo98},
K80 = \citet{kuh80},
L70 = \citet{lyn70},
L72 = \citet{lyn72},
O82 = \citet{oke82},
V96 = \citet{ver96},
V99 = \citet{vil99}, and
M98 = \citet{ma98}.\\
Luminosity distances ($D_{\rm L}$) and scales assume the adopted cosmological 
parameters, $H_{0}=71~$km~s$^{-1}$~Mpc$^{-1}$, $\Omega_{\rm M}=0.27$, and 
$\Omega_{\rm \Lambda}=0.73$.
\end{table}

\begin{table}
\scriptsize
\caption[]{\label{table-2k} VLA Data Summary}
\begin{center}
\begin{tabular}{lccccccc}
\hline \hline
Name & Program & Date & Frequency & Array & Exp.~Time & Beam$^{***}$ \\ 
     &               &       & (GHz)  &  & (s)   & (\arcsec)  \\
\hline
0730$+$257   & AK353 & 1994 Mar 20 & 8.7  & A & 2440 & 0.35 \\
0805$+$046   & AB560 & 1990 Mar 24 & 4.9  & A & 2420 & 0.4  \\
1311--270  & AL119 & 1986 Apr 28 & 4.9 & A & 1210 & 0.7  \\ 
1318+113   & AB322 & 1985 Mar 10 & 4.9  & A & 1590 & 0.25 \\  
1834+612   & AT165 & 1994 Sep 06 & 4.7  & B &  170 & 1.5  \\ 
0833$+$585   & AL164 & 1987 Oct 09 & 4.9  & A & 5200 & 0.50 \\
1239+376   & AT165 & 1994 Sep 06 & 4.7  & B &  170 & 4.2, 1.4 at PA = $-64.5\deg$ \\
1754+676   & AC755 & 2004 Dec 04 & 1.4  & A & 1060 & 1.5, 1.1 at PA = $19.7\deg$\\
1418--064  & S8723 & 2007 Jul 27 & 4.9  & A & 3570 & 0.66, 0.38 at PA = $-20.7\deg$ \\
1428+422*  & AC755 & 2004 Dec 06 & 1.4  & A & 2020 & 1.6, 1.1 at PA = $52.7\deg$ \\
1508+5714**    & AM492 & 1995 Jul 14 & 1.4  & A & 300 & 1.5, 1.0 at PA = $-4.4\deg$ \\
\hline \hline
\end{tabular}
\end{center}
Notes -- $^{*}$ The data for 1428+422 were published in \citet{che12}.
$^{**}$ The data for 1508+5714 were published in \citet{che04}.
$^{***}$ The beam sizes are elliptical Gaussians with major axis ($"$), minor axis ($"$), at the position angles (PA) indicated, or circular Gaussians when a single dimension is indicated.
\end{table}

\begin{table}
\caption[]{\label{table-chandra} \Chandra\ X-ray Observation Summary}
\begin{center}
\begin{tabular}{lccc}
\hline \hline
Name    & ObsID & Date & Net Exposure \\
        &       &      & (ksec)       \\
\hline
0730$+$257   & 10307 & 2009 Feb 12 & 20.1  \\
0805$+$046   & 10308 & 2009 Feb 20 & 19.2  \\
1311$-$270  & 10309 & 2009 Mar 19 & 18.3  \\
1318$+$113   & 10310 & 2009 Mar 05 & 18.3  \\
1834$+$612   & 10311 & 2009 May 07 & 17.2  \\
0833$+$585   & 7870  & 2007 Jan 12 & 3.8  \\
1239$+$376   & 7871  & 2007 Mar 10 & 4.7 \\
1754$+$676   & 7872  & 2007 May 25 & 6.5 \\    
1418$-$064  & 7873  & 2007 Jun 04 & 3.3 \\     
1428$+$422*  & 7874  & 2007 Mar 26 & 10.6 \\    
1508$+$5714** & 2241  & 2001 Jun 10 & 88.9 \\
\hline \hline
\end{tabular}
\end{center}
Notes -- 
$^{*}$ The data for 1428$+$422 were published in \citet{che12}.
$^{**}$ The data for 1508$+$5714 were published in \citet{sie03}.
\end{table}

\begin{sidewaystable*}
{\tiny{
\centering
\vspace{0.4cm}
\caption{Model Parameters for Quasars}
\label{tbl:quasar2}
\begin{tabular}{lc|ccccc|cccccc}
\hline\hline
\\
Source & obsID & \multicolumn{5}{c}{{\it{Power Law with Galactic Absorption Model}}}  & \multicolumn{6}{c}{{\it{Power Law with the Intrinsic Absorption Model}}}\\
\hline
& & Norm & $\Gamma_X$ & f(0.5-2 keV) & f(2-10 keV) & stat & Norm  & $\Gamma_X$ & $N_H(z_{qso})$ & f(0.5-2 keV) & f(2-10 keV)& stat \\
  & (1) &  (2) & (3) & (4) & (5) & (6) &(2) & (3) & (7) & (4) & (5) & (6) \\
\\
\hline
\\
0730+257& 10307
&  $ 10.70^{+1.14}_{-1.07}$ & $ 1.61^{+0.14}_{-0.14}$ & $ 2.43^{+0.25}_{-0.28}$ & $ 4.92^{+1.39}_{-0.99}$ & 275.17 & $ 13.27^{+3.13}_{-2.50}$ & $ 1.82^{+0.23}_{-0.22}$ & $ 1.83^{+1.67}_{-1.54}$  & $ 2.97^{+0.65}_{-0.60}$ & $ 4.62^{+1.93}_{-1.53}$ & 273.73 \\
0805+046& 10308
& $ 21.07^{+1.55}_{-1.48}$ & $ 1.71^{+0.10}_{-0.10}$ & $ 4.73^{+0.36}_{-0.34}$ & $ 8.53^{+1.50}_{-1.31}$ & 306.77 & $ 23.61^{+3.69}_{-3.18}$ & $ 1.82^{+0.17}_{-0.16}$ & $ < 2.80$ & $ 5.29^{+0.80}_{-0.74}$ & $ 7.92^{+2.87}_{-1.99}$ & 305.95 \\
1311--270&10309
& $ 73.35^{+3.08}_{-2.99}$ & $ 1.79^{+0.06}_{-0.06}$ & $ 16.37^{+0.61}_{-0.62}$ & $ 26.19^{+2.94}_{-2.3}$ & 382.97 & $ 78.03^{+6.91}_{-6.30}$ & $ 1.85{\pm0.10}$ & $ <1.27$ & $ 17.36^{+1.46}_{-1.42}$ & $ 25.08^{+4.57}_{-4.15}$ & 382.26 \\
1318+113 & 10310
&  $ 20.08^{+1.49}_{-1.43}$ & $ 1.82^{+0.11}_{-0.11}$ & $ 4.53^{+0.34}_{-0.35}$ & $ 6.77^{+1.31}_{-1.07}$ & 273.94 & $ 25.40^{+4.33}_{-3.68}$ & $ 2.07{\pm0.19}$ &  $ 1.22^{+0.76}_{-0.73} $  & $ 5.72^{+0.85}_{-0.89}$ & $ 5.96^{+2.06}_{-1.59}$ & 271.05 \\
1834+612 & 10311
&  $ 186.33^{+4.93}_{-4.86}$ & $ 1.49^{+0.03}_{-0.03}$ & $ 42.52^{+1.14}_{-1.16}$ & $ 106.77^{+6.23}_{-6.51}$ & 478.76 & $ 206.95^{+11.03}_{-10.44}$ & $ 1.58{\pm0.05}$ & $ 0.65^{+0.28}_{-0.28}$ & $ 46.93^{+2.27}_{-2.34}$ & $ 100.74^{+11.9}_{-9.94}$ & 473.15 \\
0833+585  &7870
&  $ 185.14^{+10.32}_{-10.00}$ & $ 1.42^{+0.07}_{-0.07}$ & $ 42.41\pm2.3$ & $ 118.89^{+16.31}_{-14.85}$ & 417.16 & $ 185.18^{+15.38}_{-10.06}$ & $ 1.42^{+0.09}_{-0.07}$ & $ < 0.34$  & $ 42.29^{+2.5}_{-2.43}$ & $ 118.71^{+14.4}_{-15.06}$ & 417.16 \\
1239+376 & 7871
&   $ 27.76^{+3.46}_{-3.20}$ & $ 1.45^{+0.16}_{-0.16}$ & $ 6.40^{+0.73}_{-0.80}$ & $ 16.75^{+5.15}_{-3.92}$ & 249.68 & $ 30.62^{+8.30}_{-5.69}$ & $ 1.54^{+0.26}_{-0.23}$ & $ < 11.8 $ & $ 6.82^{+1.32}_{-1.25}$ & $ 15.53^{+6.54}_{-5.15}$ & 249.46 \\
1754+676 & 7872
&  $ 24.29^{+2.91}_{-2.69}$ & $ 1.67^{+0.16}_{-0.16}$ & $ 5.51^{+0.62}_{-0.60}$ & $ 10.42^{+3.48}_{-2.71}$ & 238.03 & $ 24.28^{+4.72}_{-2.69}$ & $ 1.67^{+0.21}_{-0.16}$ & $ < 8.2 $ & $ 5.45^{+0.65}_{-0.82}$ & $ 10.13^{+4.16}_{-2.62}$ & 238.03 \\
1418--064 &7873
&   $ 27.12^{+4.22}_{-3.84}$ & $ 1.74^{+0.22}_{-0.21}$ & $ 6.11\pm0.89$ & $ 10.43^{+4.35}_{-3.16}$ & 176.78 & $ 27.12^{+4.38}_{-3.84}$ & $ 1.74^{+0.22}_{-0.21}$ & $ < 7.76 $ & $ 6.10^{+0.93}_{-0.88}$ & $ 10.46^{+4.98}_{-3.21}$ & 176.78 \\
1428+422 & 7874
& $ 239.52^{+6.61}_{-6.51}$ & $ 1.36^{+0.04}_{-0.04}$ & $ 55.39^{+1.48}_{-1.46}$ & $ 169.50^{+9.81}_{-11.62}$ & 528.73 & $ 262.72^{+15.48}_{-14.54}$ & $1.44{\pm0.06}$ & $ 2.32^{+1.29}_{-1.26}$ & $ 60.68^{+3.19}_{-3.63}$ & $ 162.72^{+16.92}_{-16.52}$ & 525.29 \\
1508+5714 & 2241
& $ 64.36^{+1.11}_{-1.10}$ & $ 1.55^{+0.02}_{-0.02}$ & $ 14.60^{+0.23}_{-0.24}$ & $ 33.53^{+1.49}_{-1.39}$ & 522.54 & $ 65.50^{+2.31}_{-2.03}$ & $ 1.56{\pm0.04}$ & $ < 1.73 $ & $ 14.80^{+0.54}_{-0.53}$ & $ 33.35^{+2.41}_{-2.30}$ & 522.20 \\
\\
\hline\hline
\end{tabular}
\\
Notes -- (1) \Chandra\ Obsid -- (2) Norm - power law normalization in
[$10^{-6}$] photons~cm$^{-2}$~s$^{-1}$. -- (4, 5) Fluxes [$10^{-14}$]
erg~cm$^{-2}$~s$^{-1}$ -- (6) the {\tt cstat} value calculated in
Sherpa -- (7) intrinsic absorption at the quasar redshift -- The
assumed model in {\it Sherpa}: {\tt
xsphabs(Gal)*xszphabs(N($z_{qso}$))*powlaw1d}, where Gal is the
equivalent Hydrogen column density in the Milky Way from {\it
Colden}(Stark et al. 1992) and it is fixed in the model;
N$_H$($z_{qso}$) is the column density at the quasar redshift which is
a fit parameter. -- The fluxes in the table are corrected for
absorption and are calculated in the observer frame (i.e. not
k-corrected).\\}}
\end{sidewaystable*}





\begin{table*}
{\scriptsize{
\caption{Parameters for each region.}
\label{tbl:sigx}
\centering
\begin{tabular}{llccccc}
\hline \hline
ObsID   & Region$^a$ & Jet X-Ray Flux$^b$ & Radio Flux &p-value&Signif.$^c$ &Avg.\\
        &       & $10^{-15}$(erg/s/cm$^2$)    & $10^{-15}$(erg/s/cm$^2$)  & ($\hat u$)      &($\alpha=0.02$) &$\tau_1$\\
\hline
10307 & Q       & 0.009 & 0.80  & 0.088 &           &0.82  \\
      & 1       & 0.003 & 0.76 & 0.792 &           & 0.03   \\
      & 2       & 0.622& 1.18 & 0.006 & Yes       & 5.50   \\
      & 3       & 1.98 & 4.81 & 0.005 & Yes       & 13.14   \\
      & C       &  &  & 0.221 &           & 6.39  \\
10308 & Q       & 0.023 & 0.61  & 0.132 &           & 1.16  \\
      & 1       & 0.043 & 1.37 & 0.048 &           & 0.39   \\
      & 2       & 0.140 & 16.00 & 0.018 & Marginal  & 1.00   \\
      & 3       & 0.163 &0.17 & 0.020 & Marginal  & 0.77   \\
      & C       & & &  0.075 &           & 5.61  \\
10309 & Q       & 0.041 & 5.28 &  0.156 &           & 0.64  \\
      & 1       & 0.004 & 0.43 &  1.000 &           & 0.02   \\
      & 2       &  1.48 & 0.35 &  0.005 & Yes       & 8.53   \\
      & 3       & 0.347 & 0.38 &  0.008 & Yes       & 1.93   \\
      & 4       &  0.180 &1.52&  0.009 &  Yes      & 1.27   \\
      & 5       &  3.22 &2.71 &  0.005 &  Yes      & 17.02   \\
      & C       &  & &  0.452 &           & 5.05 \\
10310 & Q       & 0.046& 0.36 &  0.238 &           &0.15  \\
      & 1       & 0.219 & 6.06  &  0.018 &  Marginal & 1.24   \\
      & 2       & 0.022 & 4.52 &  0.432 &           & 0.11   \\
      & 3       & 1.43 &16.00 &  0.005 & Yes       & 8.62   \\
      & 4       & 0.196 & 11.9&  0.018 &  Marginal & 1.09   \\
      & C       &  &  & 0.036 &           & 8.09 \\
10311 & Q       & 9.63 & 20.50& 0.005 & Yes       &51.54  \\
      & 1       & 0.129 & 0.930 &  0.028 &           & 0.67   \\
      & 2       & 0.486 & 0.0091 &  0.008 & Yes       & 2.79   \\
      & C       & &  & 0.018 & Marginal  & 13.10 \\
 7870 & Q       &0.383 & 35.00 & 0.068 &           & 0.80  \\
      & 1       &0.964 & 0.26 & 0.011 & Marginal  & 1.10   \\
      & 2       &8.05 &0.10 & 0.005 & Yes       & 9.82   \\
      & 3       &6.88 & 0.15 & 0.005 & Yes       & 8.54   \\
      & C       & & & 0.057 &           & 6.91  \\
 7871 & Q       & 0.457 & 12.60 &  0.528 &           &0.81  \\
      & 1       & 0.059 & 0.06&  0.950 &           & 0.11   \\
      & C       & &    &  1.000 &           & 2.41 \\
 7872 & Q       &  0.143 & 0.75  &  0.731 &           &0.46  \\
      & 1       &  0.447 & 0.03 &  0.792 &           & 0.90   \\
      & C       & & & 0.559 &           & 2.86 \\
 7873 & Q       & 1.15 &10.70 &  0.058 &           &2.24  \\
      & 1       & 15.6 & 0.03 &  0.005 & Yes       & 17.19   \\
      & C       & & & 0.528 &           & 1.83 \\     
 7874 & Q       & 20.19 &2.20  & 0.005 & Yes       &70.01  \\
      & 1       & 1.23 & 0.02 &  0.010 & Marginal  & 4.26   \\
      & C       & & & 1.000 &           & 1.79 \\
 2241 & Q       & 2.51& 2.54 & 0.005  & Yes       & 82.50  \\
      & 1       & 3.25&0.02  & 0.005  & Yes       & 106.84   \\
      & C       & & & 0.005  & Yes       & 53.12  \\
   \hline
   \hline
\end{tabular}

NOTES:

$^a:$ Q - excess emission in the quasar core region, excluding the quasar; $1-5$ - numbering of identified jet regions based on radio data;
C - a region complementary to the quasar and a jet (the entire image minus the quasar core and jet regions)

$^b:$ Jet X-Ray Flux is the Flux calculated using the multiscale counts, so the background and quasar are removed

$^c:$ - 'Yes' marks a significant detection $\alpha <0.01$; 'Marginal' marks a detection with $0.01< \alpha <0.02$}}
\end{table*}



\begin{table*}
\begin{center}
\caption{X-ray Spectral Fits for selected Jets Regions}
\label{tbl:jets}
\begin{tabular}{lcccccccc}
\hline\hline
\\
source & obsID & region & Norm & $\Gamma_X$ & f(0.5-2 keV) & f(2-10 keV) & stat \\
 & &  (1) & (2) & (3) & (4) & (5) & (6) \\
\\
\hline
\\
0730+257& 10307 & 3 & $ 1.07^{+0.39}_{-0.31}$ & $ 1.93^{+0.51}_{-0.48}$ & $ 0.24^{+0.08}_{-0.09}$ & $ 0.31^{+0.32}_{-0.17}$ & 69.77 \\
& 10307 & 2 & $ 0.97^{+0.37}_{-0.30}$ & $ 1.67^{+0.49}_{-0.47}$ & $ 0.22^{+0.07}_{-0.08}$ & $ 0.38^{+0.55}_{-0.22}$ & 78.06 \\
0805+046 &10308 & 2 & $ 0.51^{+0.27}_{-0.20}$ & $ 2.04^{+0.77}_{-0.71}$ & $ 0.12^{+0.05}_{-0.05}$ & $ 0.13^{+0.29}_{-0.09}$ & 46.57 \\
1311--270 & 10309 &  5  & $ 2.21^{+0.57}_{-0.49}$ & $ 2.82^{+0.51}_{-0.48}$ & $ 0.53^{+0.12}_{-0.12}$ & $ 0.17^{+0.20}_{-0.08}$ & 75.52 \\
& 10309 & 2  & $ 0.91^{+0.38}_{-0.30}$ & $ 1.48^{+0.49}_{-0.48}$ & $ 0.21^{+0.08}_{-0.08}$ & $ 0.51^{+0.66}_{-0.3}$ & 69.11 \\
1318+113 & 10310 & 2 & $ 0.91^{+0.34}_{-0.27}$ & $ 1.92^{+0.55}_{-0.52}$ & $ 0.20^{+0.08}_{-0.08}$ & $ 0.26^{+0.37}_{-0.15}$ & 61.89 \\
1834+612 & 10311 & 2 & $ 0.55^{+0.31}_{-0.22}$ & $ 1.05^{+0.55}_{-0.54}$ & $ 0.14^{+0.05}_{-0.06}$ & $ 0.58^{+1.2}_{-0.38}$ & 65.61 \\
0833+585& 7870 &  2 & $ 3.09^{+1.50}_{-1.12}$ & $ 0.94^{+0.47}_{-0.46}$ & $ 0.76^{+0.32}_{-0.33}$ & $ 3.79^{+6.35}_{-2.13}$ & 76.34 \\
& 7870 &  3 & $ 5.02^{+1.84}_{-1.49}$ & $ 1.73^{+0.49}_{-0.47}$ & $ 1.13^{+0.37}_{-0.39}$ & $ 1.92^{+2.6}_{-1.02}$ & 77.39 \\
& 7870 &  1 & $ 0.46^{+0.67}_{-0.34}$ & $ 3.93^{+3.10}_{-2.25}$ & $ 0.16^{+0.25}_{-0.05}$ & $ 0.01^{+0.04}_{-0.01}$ & 8.27 \\
1239+376 & 7871 & 1 & $ 0.27^{+0.48}_{-0.23}$  & $ 4.73^{+3.78}_{-2.59}$ & $ 0.10^{+0.20}_{-0.03}$ & $ <0.10 $ & 7.79 \\
1418--064 & 7873 & 1  & $ 9.96^{+2.64}_{-2.26}$ & $ 1.64^{+0.35}_{-0.34}$ & $ 2.32^{+0.5}_{-0.61}$ & $ 4.33^{+3.34}_{-1.82}$ & 116.01 \\
1428+422 & 7874 & 1 & $ 2.39^{+0.71}_{-0.59}$ & $ 1.64^{+0.39}_{-0.38}$ & $ 0.56^{+0.13}_{-0.15}$ & $ 1.07^{+0.94}_{-0.53}$ & 95.17 \\
1508+572 & 2241 & 1 & $ 2.14^{+0.20}_{-0.19}$ & $ 1.83^{+0.15}_{-0.14}$ & $ 0.48^{+0.04}_{-0.05}$ & $ 0.71^{+0.20}_{-0.17}$ & 239.02 \\
\\
\hline\hline
\end{tabular}
\end{center}
Notes -- 
(1) Obsid and regions; (2) Norm [$10^{-6}$] photons~cm$^{-2}$~s$^{-1}$. -- 
(4, 5) Fluxes [$10^{-14}$] erg~cm$^{-2}$~s$^{-1}$; --
These are based on a simple absorbed power law fit with NH frozen at the Galactic
values.  The fluxes in the table are corrected for absorption and are calculated in the observer frame (i.e. not k-corrected).

\end{table*}

\FloatBarrier
\newpage
\appendix

%


\section{Flux Ratio Test}
We are interested in determining whether the flux ratio, $f_X / f_r$, correlates with redshift and more specifically if there is a difference in the distribution of the flux ratios of sources with higher redshifts  and those with lower redshifts. To investigate this, we split the detected and marginally detected jets into two redshift classes: low redshift $(z < 3)$ and high redshift $(z \geq 3)$ and denote the $\log_{10}$ flux ratio for the jet in region $i$ by   
$R_i=\log_{10}(f_X ^i / f_r ^i)$, for $i=1,..18$.  Of these, the first $15$ ratios correspond to regions of low redshift sources and the remaining three to regions with high redshift sources. We postulate that

\begin{align}
\begin{split}
R_i &\ind N(\mu_L ,\sigma _L ) \ \text { for low $z$, i.e., for }i=1,\ldots,15\\
R_i &\ind N(\mu_H ,\sigma _H ) \ \text { for high $z$, i.e., for } i=16,\ldots,18 \\
\end{split}
\label{eq:pr}
\end{align}
where $N(\mu, \sigma)$ denotes a normal distribution with mean $\mu$ and standard deviation $\sigma$.  We independently assume non-informative priors for the parameters in Equation~\ref{eq:pr},  i.e., $p(\mu_L ,\sigma _L^2 ) \propto 1/\sigma_L$ and $p(\mu_H ,\sigma _H ^2) \propto {1/\sigma_H}$. The distributions in Equation~\ref{eq:pr} can be viewed as a hierarchical prior  on $R=(R_1,\ldots, R_{18})$, hierarchical because it is specified in terms of parameters that are themselves fit to the data. We denote this hierarchical prior distribution by $p(R \mid \mu_H, \mu_L,\sigma_{H}, \sigma_L)$.

We are interested in whether or not the difference $\mu_{H}- \mu_L$ is greater than zero, which can be estimated via a Monte Carlo sample from the posterior distribution,
\begin{align}
p(\mu_{H}, \mu_L,\sigma_{H}, \sigma_L, R  \mid \yobs) 
\label{eq:post}
\end{align}
where, using the notation of Paper~I, $\yobs$ represents the observed data. 
A Monte Carlo sample from Equation~\ref{eq:post} can be obtained by 
iteratively sampling $p(R\mid \mu_{H}, \mu_L,\sigma_{H}, \sigma_L, \yobs)$ and $p(\mu_{H}, \mu_L,\sigma_{H}, \sigma_L  \mid R)$, i.e., via a Gibbs sampler. 
The second step is straightforward in that it involves a standard Bayesian fitting of the Gaussian distributions in Equation~\ref{eq:pr}. We now turn our attention to using LIRA to obtain a sample from $p(R\mid \mu_{H}, \mu_L,\sigma_{H}, \sigma_L, \yobs)$.

Each $R_i$ is a deterministic function of $\tau_1$, the expected count from the added component under the LIRA model in the region corresponding to $R_i$; again we are using the notation of Paper~I. Specifically,
\begin{align}
R_i&=\log_{10} \left( \frac{f_{X}^i}{f_{r}^i} \right )
=\log_{10} \left( \frac{\tau_1 \, c}{E_i  \, f_r ^i} \right )
\label{eq:R-tau1}
\end{align}
where $E_i$ is the average exposure in region $i$ and $c$ is a flux to photon counts conversion factor.  Thus, we can apply the transformation in Equation~\ref{eq:R-tau1} to the Monte Carlo sample of $\tau_1$ generated by LIRA to obtain a sample of $R_i$. 

The prior distribution used for $\tau_1$ in LIRA, however, does not correspond to the hierarchical one given in Equation~\ref{eq:pr}, but rather a gamma 
distribution\footnote{A Gamma distribution with shape parameter $a$ and rate parameter $b$ has probability density function ${\rm pdf}(x) = \frac{b^a}{\Gamma(a)} x^{a-1} e^{-bx}$, mean $a/b$, and standard deviation $\sqrt{a}/b$.}, 
$\tau_1 \sim \pi_i \, \Gam(a=1,b=20)$, where $\pi_i$ is the proportion of the image pixels that are in region $i$. Applying the transformation in Equation~\ref{eq:R-tau1} to this prior distribution yields the prior implicitly  assumed by LIRA for $R_i$,
\begin{equation}
R_i \ind \log_{e} \Gam\left(a=1,b=\frac{c \ \pi_i}{20 \ E_i \  f_r ^i}\right).
\label{eq:priorR}
\end{equation}
We denote this prior distribution by $p\lira(R)= \prod_{i=1}^{18}p\lira(R_i)$. The difference between $p\lira(R)$ and $p(R)$ means that LIRA produce a Monte Carlo sample from 

\begin{eqnarray}
p\lira(R|\yobs) &\propto& \frac{p(R\mid \mu_{H}, \mu_L,\sigma_{H}, \sigma_L, \yobs)  \ p\lira(R) }{p(R \mid \mu_{H}, \mu_L,\sigma_{H}, \sigma_L)}. \nonumber
\end{eqnarray}

 To derive this expression, note that more precisely, LIRA provides a Monte Carlo sample from the joint posterior distribution, $p(\theta, R \mid \yobs)$, where $\theta$ represents a number of other unknown parameters that are not pertinent to the current discussion. Thus, 
\small
\begin{eqnarray}
p\lira(R|\yobs) &=& \int p\lira(\theta, R \mid \yobs) {\rm d}\theta \\
&=&
\int {p(\yobs,\mid \theta, R) \ p(\theta | R) \ p\lira(R) \over p\lira (\yobs)} \ {\rm d}\theta \nonumber\\
&=&
{ p\lira(R) \over p\lira (\yobs)}
\int {p(\yobs,\mid \theta, R) \ p(\theta | R)} \ {\rm d}\theta, \label{eq:der-lira}
\end{eqnarray}

where 
$$p\lira (\yobs) = \int p(\yobs,\mid \theta, R) \, p(\theta | R) \, p\lira(R) \, {\rm d}\theta{\rm d}R.$$
 Because $\yobs$ and $(\mu_{H}, \mu_L,\sigma_{H}, \sigma_L)$ are conditionally independent given $R$, we can write our target posterior distribution 
\begin{eqnarray}
p(R  \mid \mu_{H}, \mu_L,\sigma_{H}, \sigma_L, \yobs) 
&=& \int p(\theta \mid R, \yobs) \ p(R  \mid \mu_{H}, \mu_L,\sigma_{H}, \sigma_L, \yobs) \ {\rm d}\theta
\nonumber \\
&=&
\int {p(\yobs,\mid \theta, R) \ p(\theta | R) \ p(R \mid \mu_{H}, \mu_L,\sigma_{H}, \sigma_L) \over p (\yobs)} \ {\rm d}\theta \nonumber \\
&=&
{p(R \mid \mu_{H}, \mu_L,\sigma_{H}, \sigma_L), \over p (\yobs)} 
\int p(\yobs,\mid \theta, R) \  p(\theta | R) \ {\rm d}\theta,
 \label{eq:der-target}
\end{eqnarray}
where $$p(\yobs) = \int p(\yobs,\mid \theta, R) \, p(\theta | R) \, p(R \mid \mu_{H}, \mu_L,\sigma_{H}, \sigma_L) \, {\rm d}\theta{\rm d}R.$$  Finally combining Equations~\ref{eq:der-lira} and \ref{eq:der-target}, we have
\begin{eqnarray}
p\lira(R \mid \yobs) &=& {
p(R  \mid \mu_{H}, \mu_L,\sigma_{H}, \sigma_L, \yobs) \ p(\yobs) \ p\lira(R)
\over
p\lira(\yobs) \ p(R \mid \mu_{H}, \mu_L,\sigma_{H}, \sigma_L)
}  \nonumber \\ 
&\propto&{
p(R  \mid \mu_{H}, \mu_L,\sigma_{H}, \sigma_L, \yobs) \ p\lira(R)
\over
p(R \mid \mu_{H}, \mu_L,\sigma_{H}, \sigma_L)
}. 
\end{eqnarray}
Our strategy is to use the LIRA posterior as a proposal rule in a  Metropolis Hastings sampler to obtain a sample from $p(R\mid \mu_{H}, \mu_L,\sigma_{H}, \sigma_L, \yobs)$.  This can be accomplished using the following algorithm. A similar strategy is employed by \cite{si2016} when they use Monte Carlo samples from the posterior distribution of the age of a number of white dwarfs obtained from separate fits as a Metropolis proposal rule for fitting a hierarchical model for all of their ages. 

\bigskip
\noindent \textbf{Sampling Algorithm}: 

\begin{description}
\item[Step 1:] Run LIRA on each of the 18 regions. For region $i$,  transform the sampled values of $\tau_1$ to $R_i$ using Equation~\ref{eq:R-tau1}. Concatenate the Monte Carlo sample of $R_1, \ldots, R_{18}$ to obtain the LIRA Monte Carlo sample of $R$. (Appropriate burn in and convergence checks should be implemented per standard practice with Markov chain Monte Carlo.)

\item[Step 2:] Set $R^{(0)}$ to a randomly selected value from the LIRA Monte Carlo sample of $R$. Fit the model in Equation~\ref{eq:pr} to $R^{(0)}$ using standard Bayesian techniques to obtain 
$\mu_L^{(0)},\mu_H^{(0)},\sigma _L ^{(0)}$, and $\sigma _H^{(0)}$.
\item[Step 3:] For $t=1,...,T$ 
\begin{description}

\item[Step 3a:] Randomly select a proposal, $R_{\text{prop}}$, from the LIRA Monte Carlo sample of $R$.

\item[Step 3b:] Compute the 18 Metropolis Hastings acceptance probabilities,

\begin{align*}
r_i&=
{p(R^{\text{prop}}_i \mid (\mu_L,\mu_H,\sigma _L , \sigma _H )^{(t-1)}, \yobs) \ p\lira(R^{(t-1)}_i \mid \yobs)
\over 
p\lira(R^{\text{prop}}_i \mid \yobs) \  p(R^{(t-1)}_i \mid (\mu_L,\mu_H,\sigma _L , \sigma _H )^{(t-1)}, \yobs)}  \nonumber \\
&=
{p(R^{\text{prop}}_i \mid (\mu_L,\mu_H,\sigma _L , \sigma _H )^{(t-1)}) \ p\lira(R^{(t-1)}_i ) 
\over p\lira(R^{\text{prop}}_i ) \  p(R^{(t-1)}_i \mid (\mu_L,\mu_H,\sigma _L , \sigma _H )^{(t-1)})} \\
&={N(R^{\text{prop}}_i; \mu^{(t-1)},\sigma^{(t-1)}) \over \log_{e} \Gamma (R^{\text{prop}}_i;a=1,b=b_i) }{\log_{10} \Gamma (R^{{(t-1)}}_i;a=1,b=b_i) \over N(R^{{(t-1)}}_i; \mu^{(t-1)},\sigma^{(t-1)})}, 
\end{align*}

where $N(R; \mu, \sigma)$ is the probability density function of a normal variable with mean $\mu$ and variance $\sigma$ evaluated at $R$, $\log_{e} \Gamma (R;a,b)$ is the probability density function of the $\log_{e}$ a gamma variable with parameters $a$ and $b$ evaluated at $R$,  $b_i={c \, \pi_i} / (20 \, E_i \, f_r ^i)$,
$\mu^{(t-1)} = \mu^{(t-1)}_L$ and $\sigma^{(t-1)} = \sigma^{(t-1)}_L$ for $i=1,\ldots, 15$, and 
$\mu^{(t-1)} = \mu^{(t-1)}_H$ and $\sigma^{(t-1)} = \sigma^{(t-1)}_H$ for $i=15,\ldots, 18$
\item[Step 3c:] For $i= 1, \ldots, 18$,  set 
$$R_i\cur = \begin{cases}
R^{\rm prop}_i  & \mbox{with probability }  \min(1, r_i) \\
R^{(t-1)}_i  & \mbox{otherwise.} \end{cases}
$$

\item[Step 3d:] Sample $\mu_L\cur,\mu_H\cur,\sigma _L \cur$, and $\sigma _H\cur$ from $(\mu_L,\mu_H,\sigma _L ,\sigma _H |R\cur) $ using standard Bayesian methods.\\
\end{description}

\item[Step 4:] A Monte Carlo estimate of the posterior probability, $\Pr(\mu_{H}- \mu_L >0|R)$ is
given by the proportion of the Monte Carlo sample for which  $\mu_{H}\cur- \mu_L\cur >0$.
\end{description}
Because this sampling algorithm in a Markov chain Monte Carlo sampler, appropriate burn in and convergence checks should be implemented.

\section{Additional Quasar Jet Observations and Analyses Figures}
As discussed in Section 3, we analyze a total of eleven quasar jets using LIRA. Here we display the radio and X-ray data as well as the results of the analyses for the remaining ten observations in the same form as Quasar 0730+257 (ObsID 10307) in Figure \ref{fig:10307}. 

\begin{figure*}[h]
\centering
\begin{tabular}[b]{@{}p{0.3\textwidth}@{}}
\centering\small (a)
\centering\includegraphics[scale=0.207,trim= 120 0 170 55,clip]{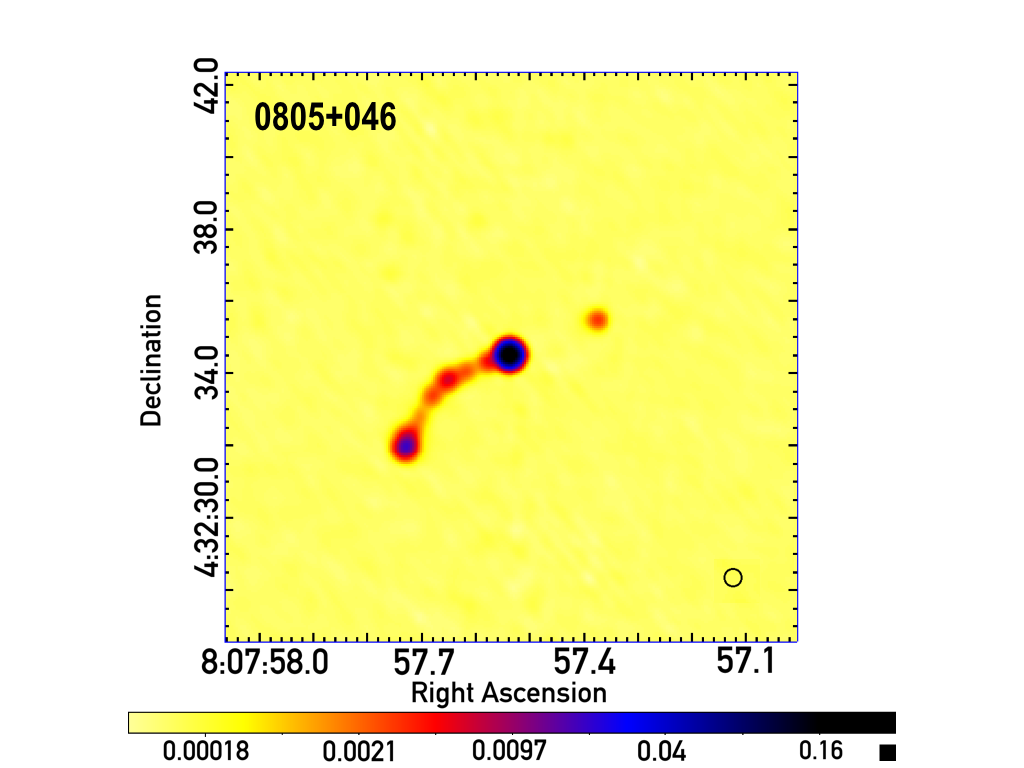}\\
\end{tabular}
\quad
\begin{tabular}[b]{@{}p{0.3\textwidth}@{}}
\centering\small (b)
\centering \includegraphics[scale=0.205,trim= 0 0 20 35,clip]{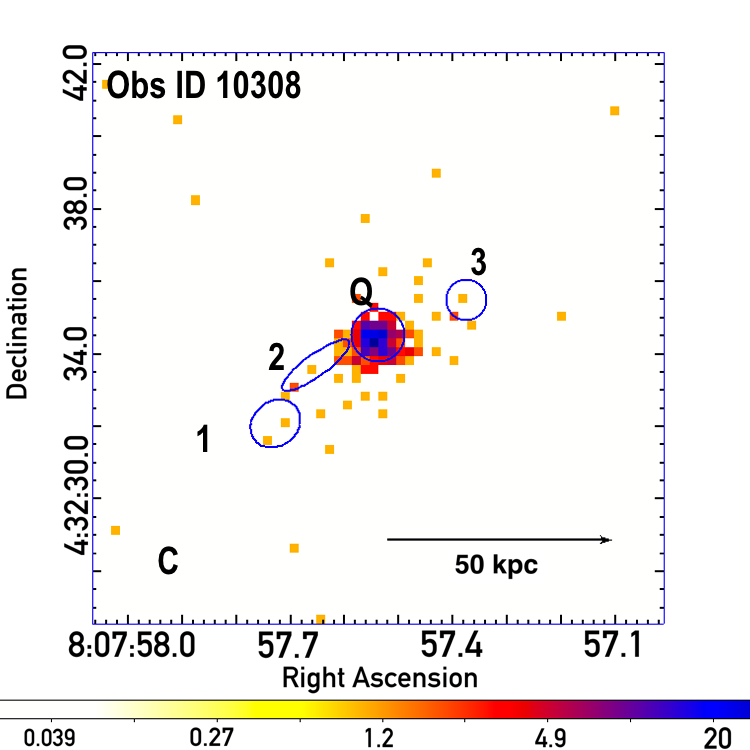}\\
\end{tabular}
\quad
\begin{tabular}[b]{@{}p{0.3\textwidth}@{}}
\centering\small (c)
\centering\includegraphics[scale=0.195,trim= 120 0 170 12,clip]{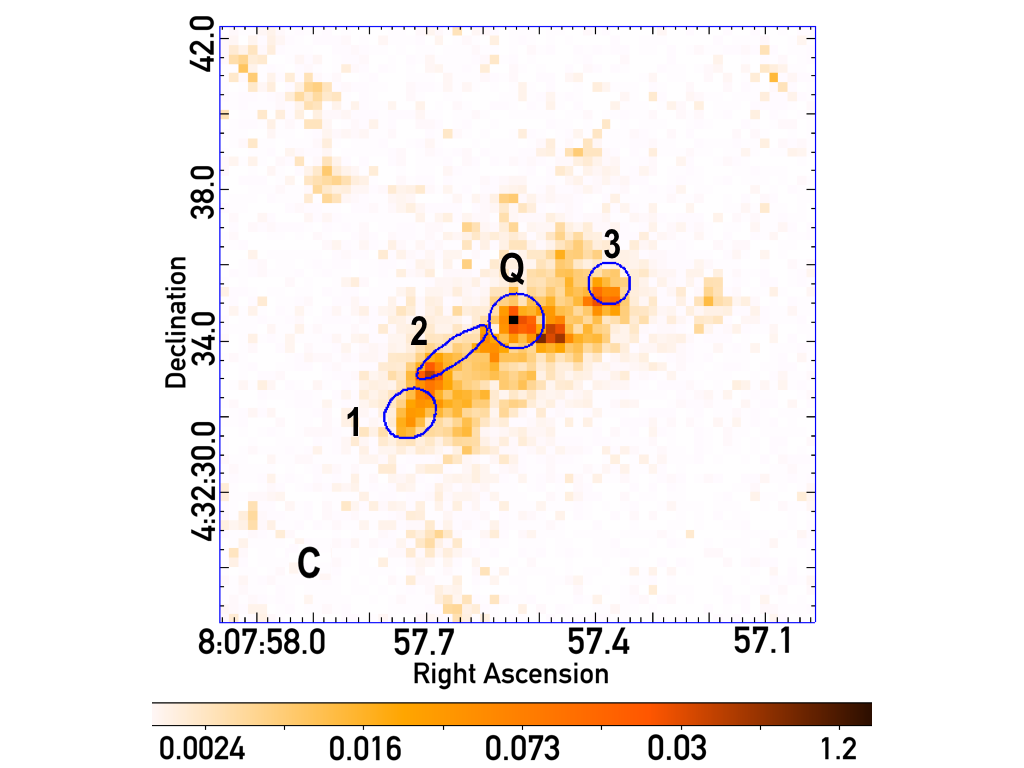}\\
\end{tabular}\\
\begin{tabular}[b]{@{}p{\textwidth}@{}}
\centering\small (d) \\
\centering\includegraphics[scale=0.5]{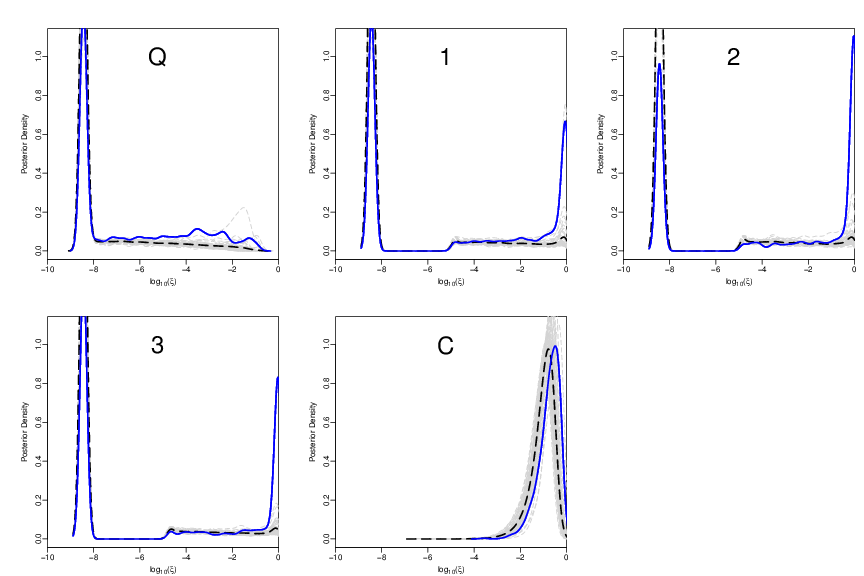}
\end{tabular}
\caption{As in Figure~\ref{fig:10307},for quasar 0805$+$046 (ObsID 10308).}
\label{fig:10308}
\end{figure*}

\begin{figure*}[h]
\centering
\begin{tabular}[b]{@{}p{0.3\textwidth}@{}}
\centering\small (a)
\centering\includegraphics[scale=0.204,trim= 120 0 170 30,clip]{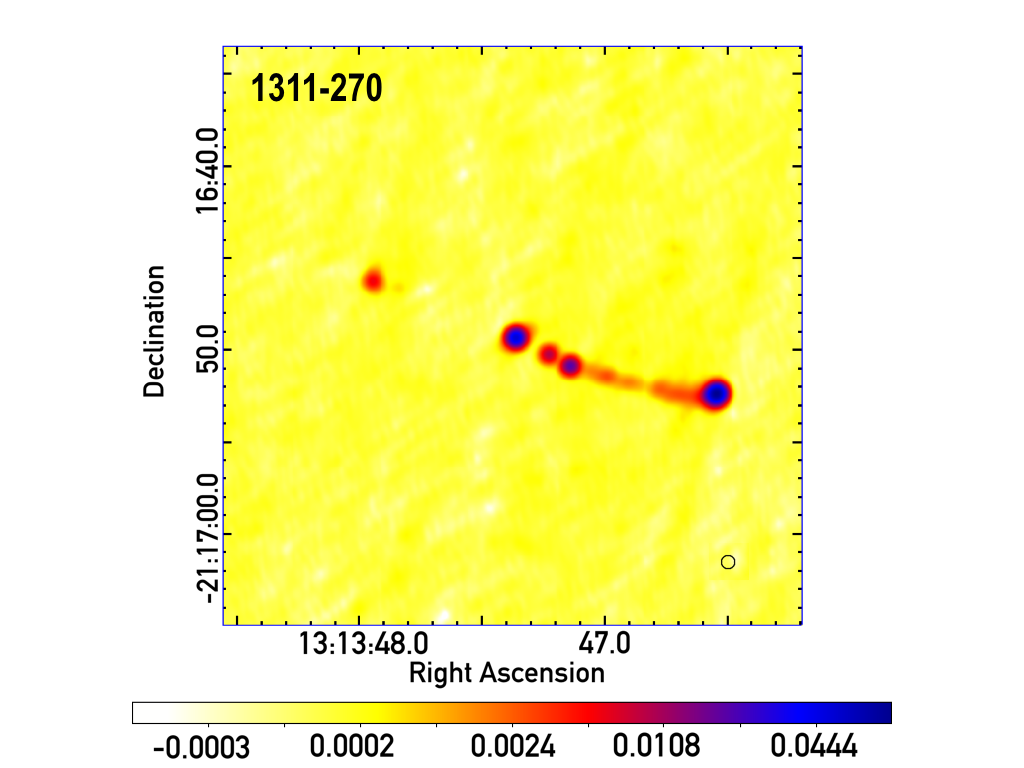}\\
\end{tabular}
\quad
\begin{tabular}[b]{@{}p{0.3\textwidth}@{}}
\centering\small (b)
\centering\includegraphics[scale=0.205,trim= 0 0 20 30,clip]{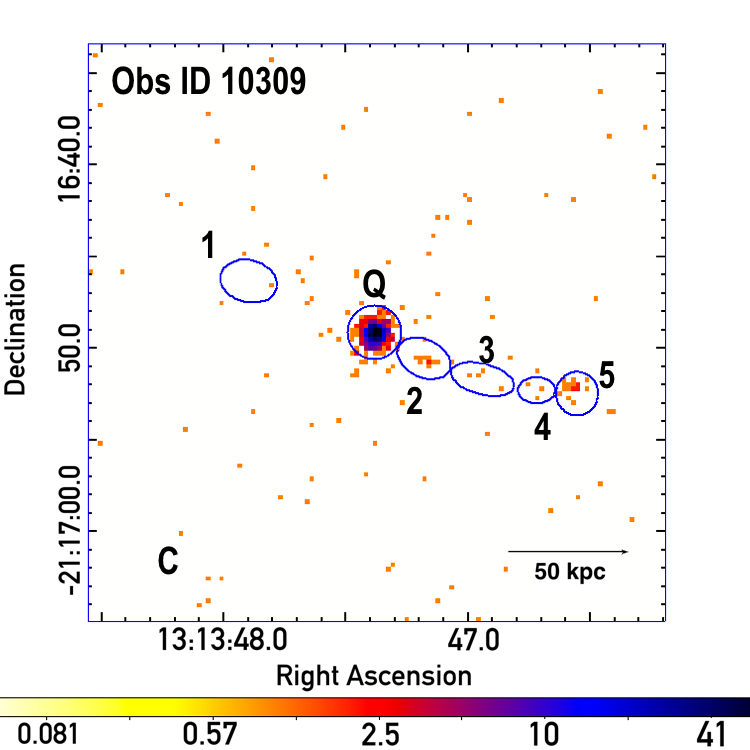}\\
\end{tabular}
\quad
\begin{tabular}[b]{@{}p{0.3\textwidth}@{}}
\centering\small (c)
\centering\includegraphics[scale=0.2,trim= 120 0 170 20,clip]{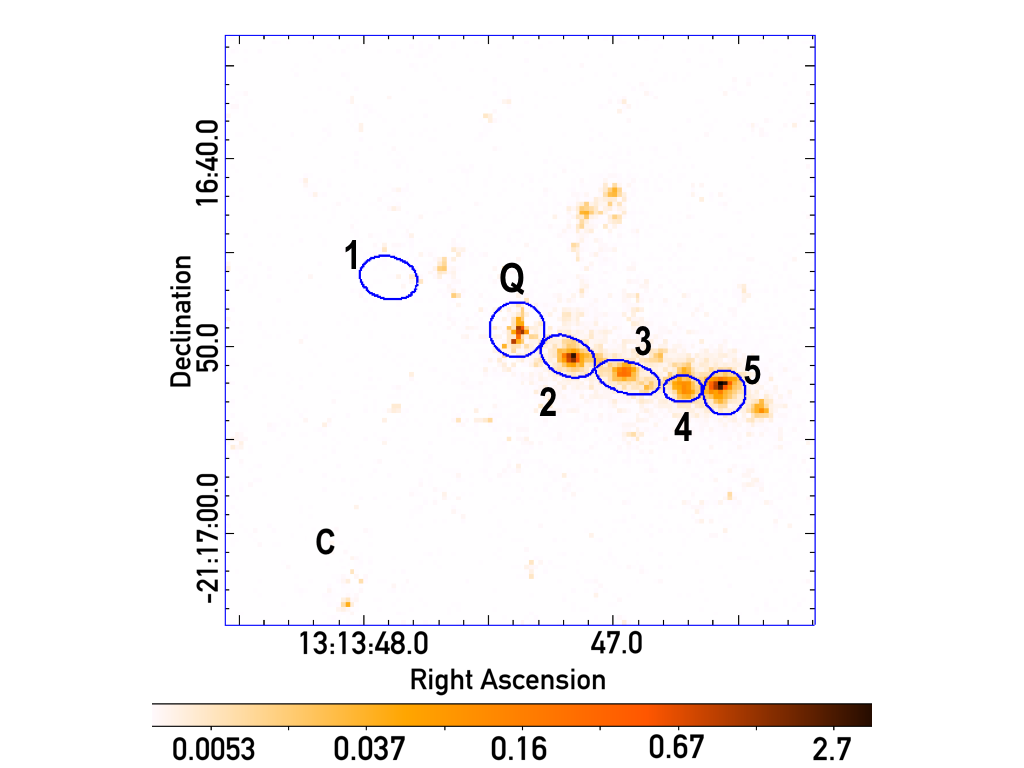}\\
\end{tabular}\\
\begin{tabular}[b]{@{}p{\textwidth}@{}}
\centering\small (d)\\
\centering\includegraphics[scale=0.5]{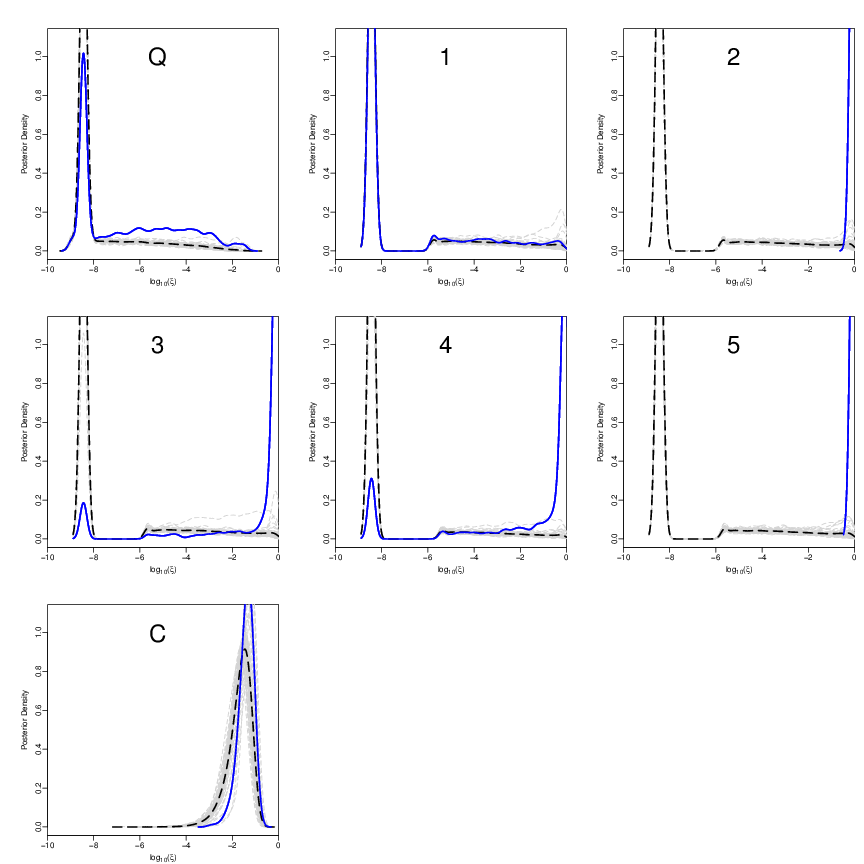}
\end{tabular}
\caption{As in Figure~\ref{fig:10307}, for quasar 1311$-$270 (ObsID 10309).}
\label{fig:10309}
\end{figure*}

\begin{figure*}[h]
\centering
\begin{tabular}[b]{@{}p{0.3\textwidth}@{}}
\centering\small (a)
\centering\includegraphics[scale=0.2,trim= 120 0 170 20,clip]{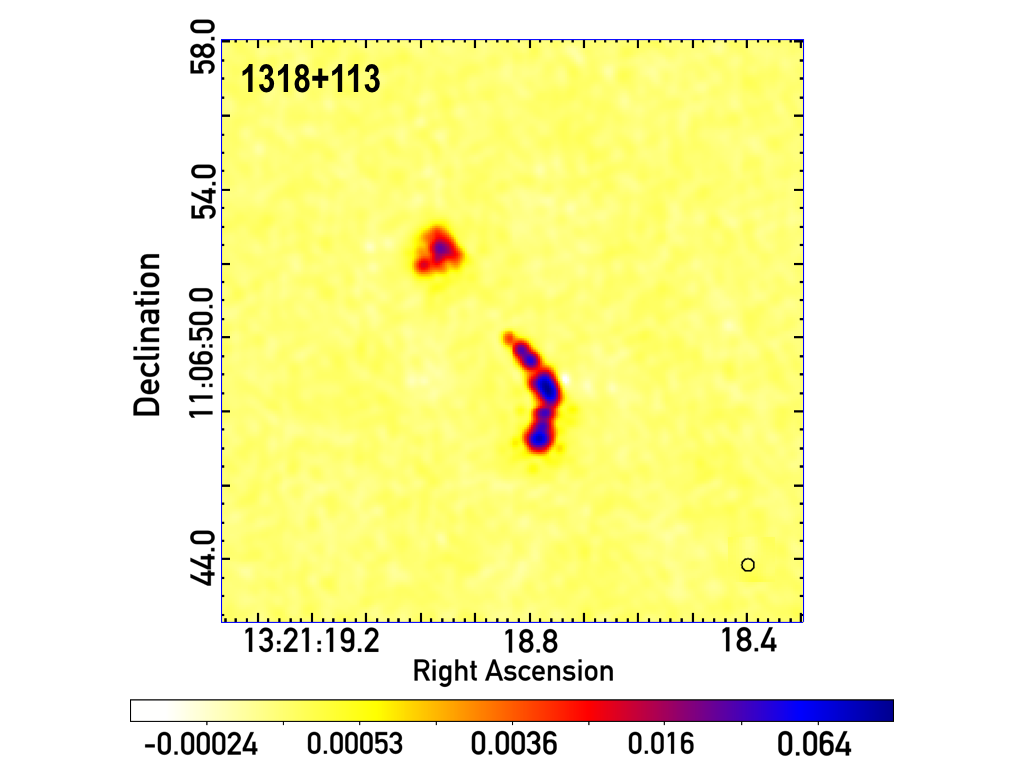}\\
\end{tabular}
\quad
\begin{tabular}[b]{@{}p{0.3\textwidth}@{}}
\centering\small (b)
\centering\includegraphics[scale=0.2,trim= 120 0 170 34,clip]{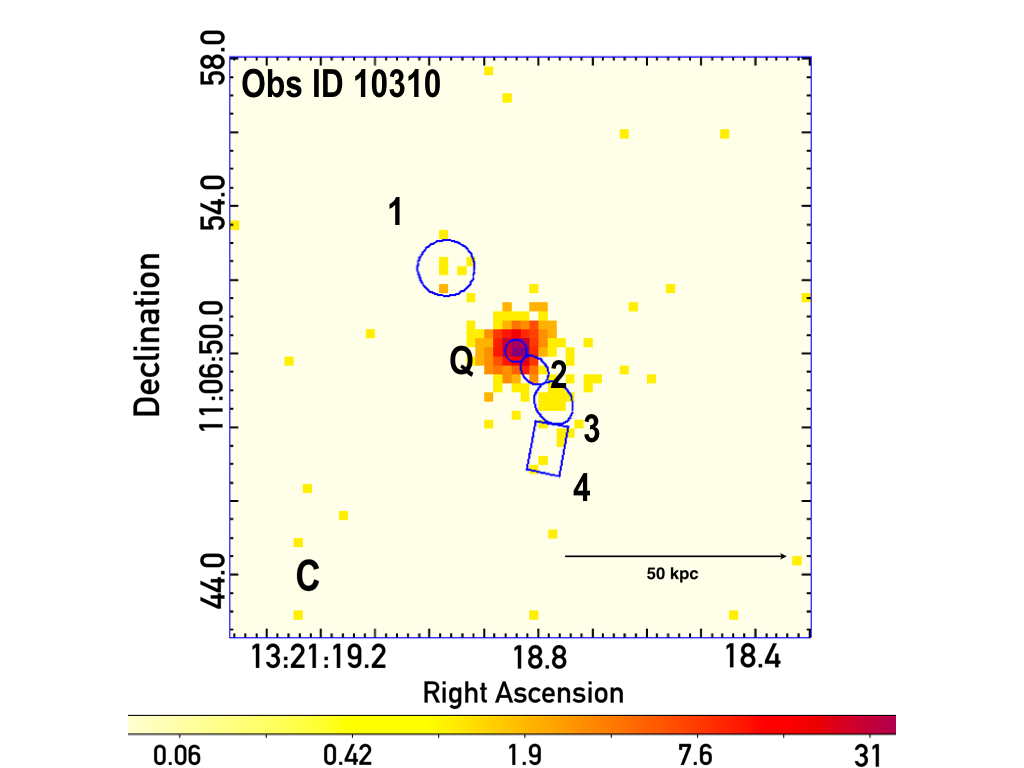}\\
\end{tabular}
\quad
\begin{tabular}[b]{@{}p{0.3\textwidth}@{}}
\centering\small (c)
\centering\includegraphics[scale=0.2,trim= 120 0 170 30,clip]{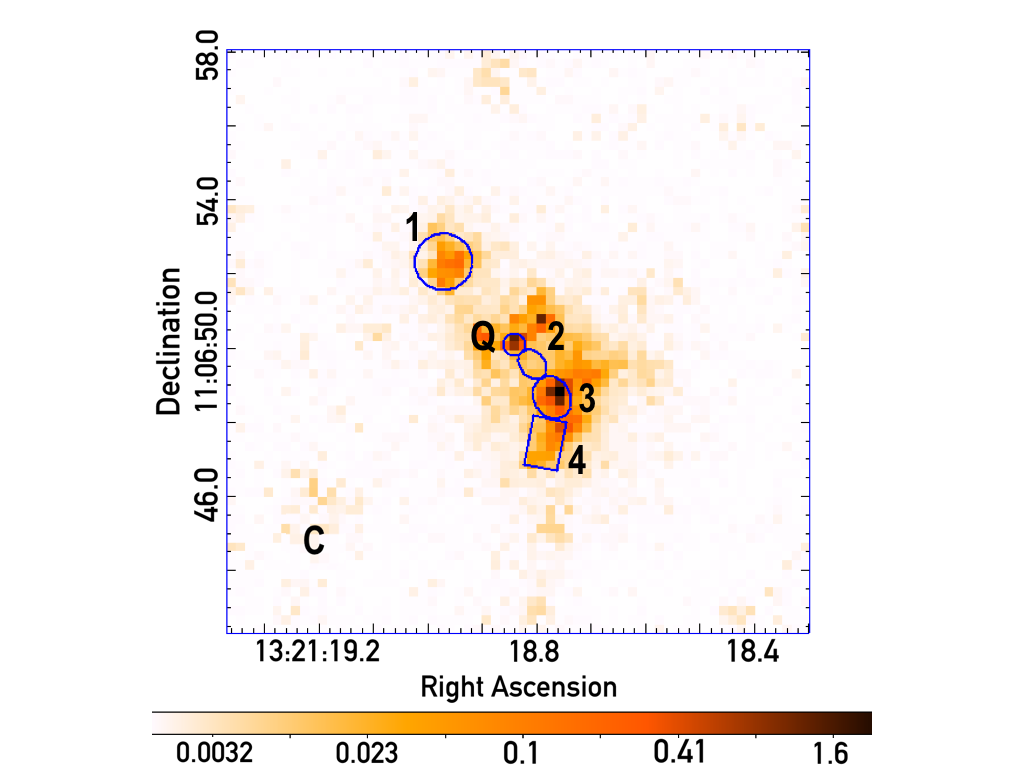}\\
\end{tabular}\\
\begin{tabular}[b]{@{}p{\textwidth}@{}}
\centering\small (d)\\
\centering\includegraphics[scale=0.5]{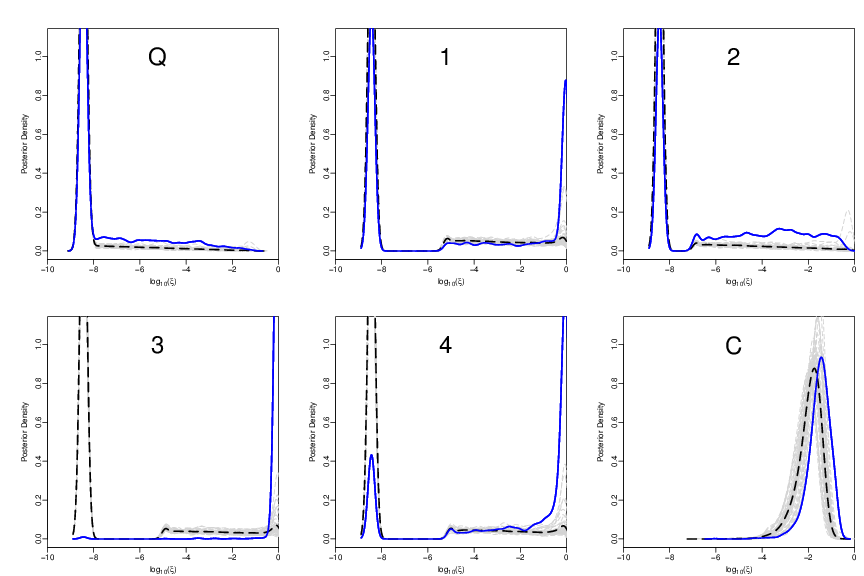}
\end{tabular}
\caption{As in Figure~\ref{fig:10307}, for quasar 1318$+$113 (ObsID 10310).}
\label{fig:10310}
\end{figure*}

\begin{figure*}[h]
\centering
\begin{tabular}[b]{@{}p{0.3\textwidth}@{}}
\centering \small (a)
\centering \includegraphics[scale=0.2,trim= 115 0 170 35,clip]{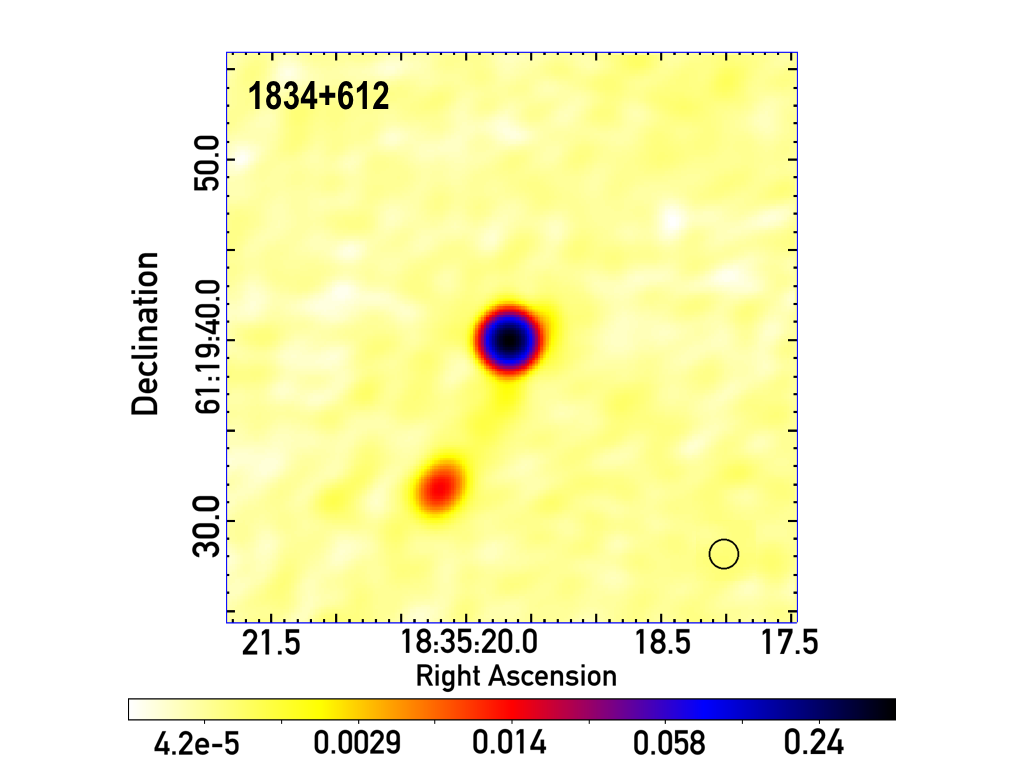}\\
\end{tabular}
\quad
\begin{tabular}[b]{@{}p{0.3\textwidth}@{}}
\centering \small (b)
\centering \includegraphics[scale=0.195,trim= 120 0 170 30,clip]{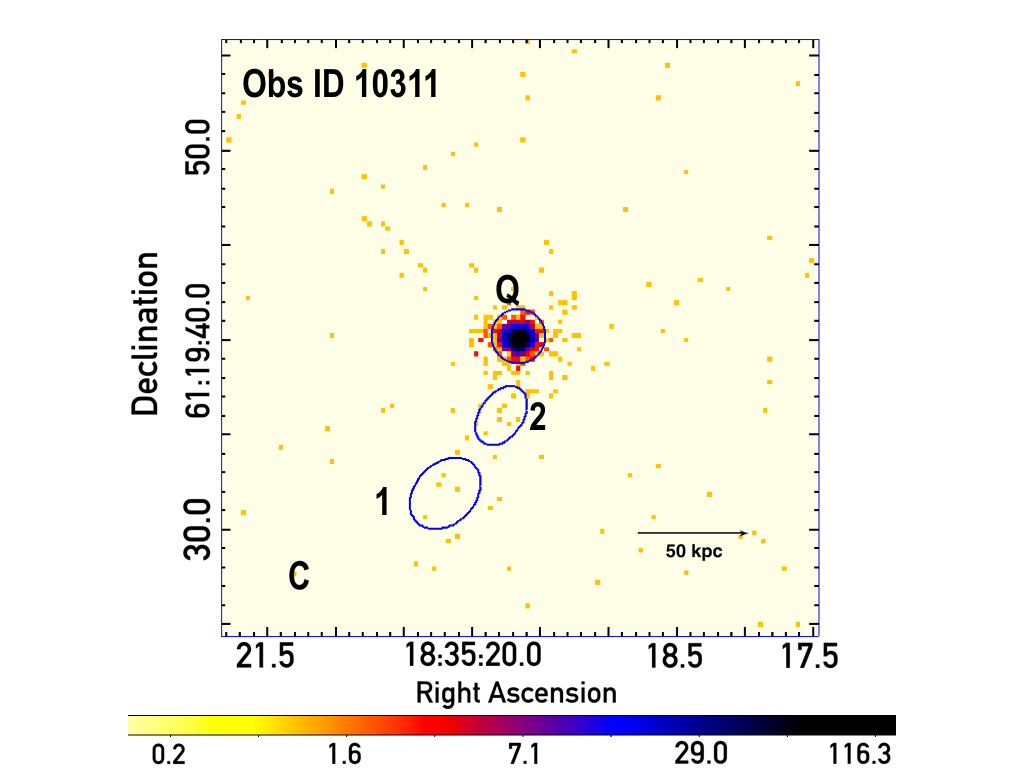}\\
\end{tabular}
\quad
\begin{tabular}[b]{@{}p{0.3\textwidth}@{}}
\centering \small (c) 
\centering\includegraphics[scale=0.2,trim= 120 0 170 30,clip]{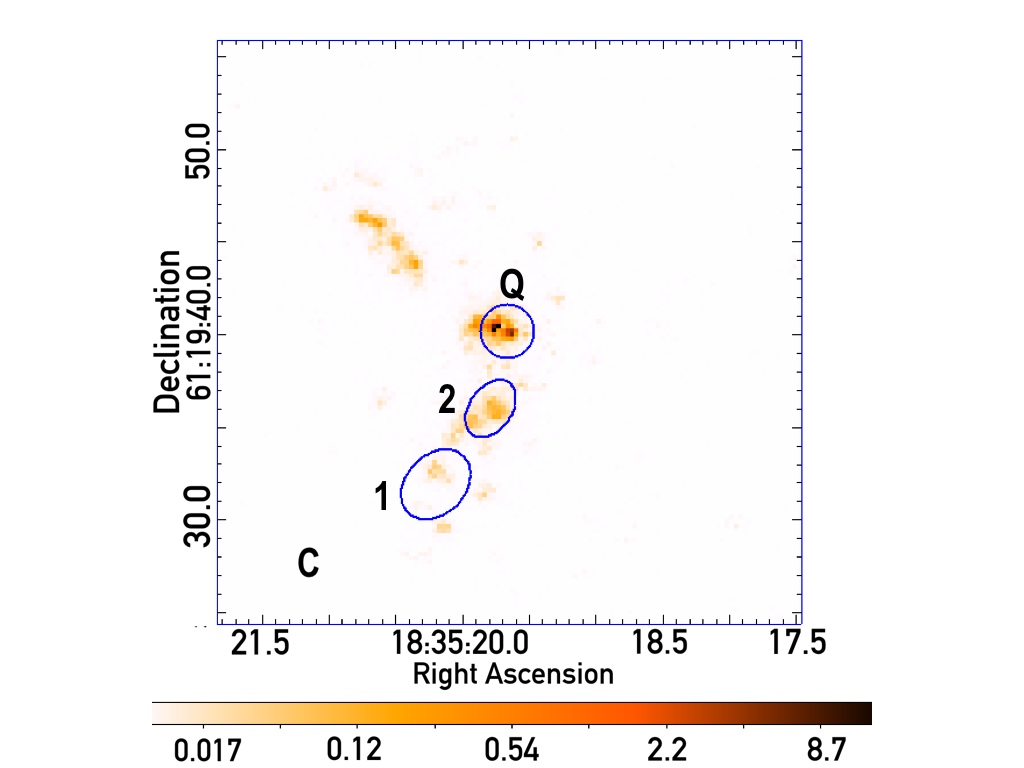}\\
\end{tabular}\\
\begin{tabular}[b]{@{}p{\textwidth}@{}}
\centering \small (d)\\
\centering \includegraphics[scale=0.5]{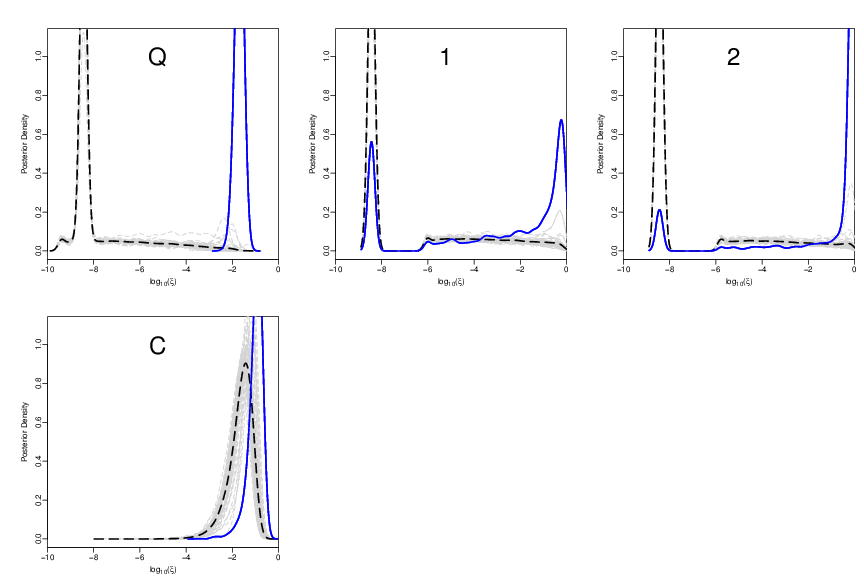}
\end{tabular}
\caption{As in Figure~\ref{fig:10307}, for quasar 1834+612 (ObsID 10311).}
\label{fig:10311}
\end{figure*}

\begin{figure*}[h]
\centering
\begin{tabular}[b]{@{}p{0.3\textwidth}@{}}
\centering \small (a)
\centering \includegraphics[scale=0.207,trim= 120 0 170 45,clip]{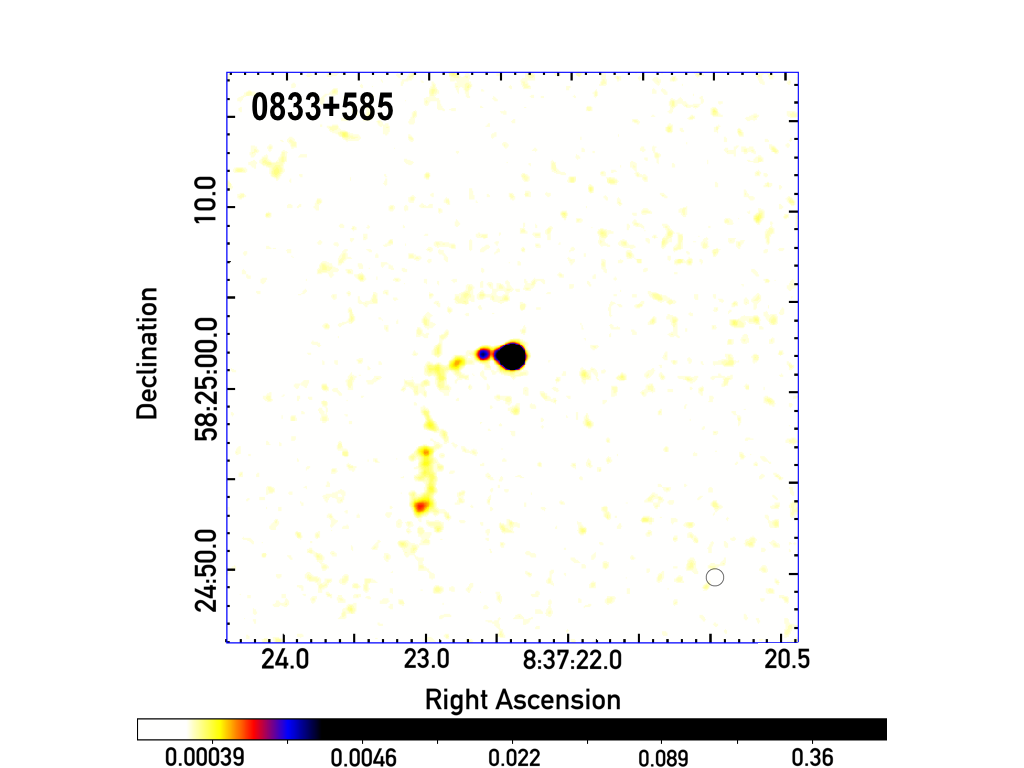} \\
\end{tabular}
\quad
\begin{tabular}[b]{@{}p{0.3\textwidth}@{}}
\centering\small (b)
\centering\includegraphics[scale=0.2,trim= 120 0 170 20,clip]{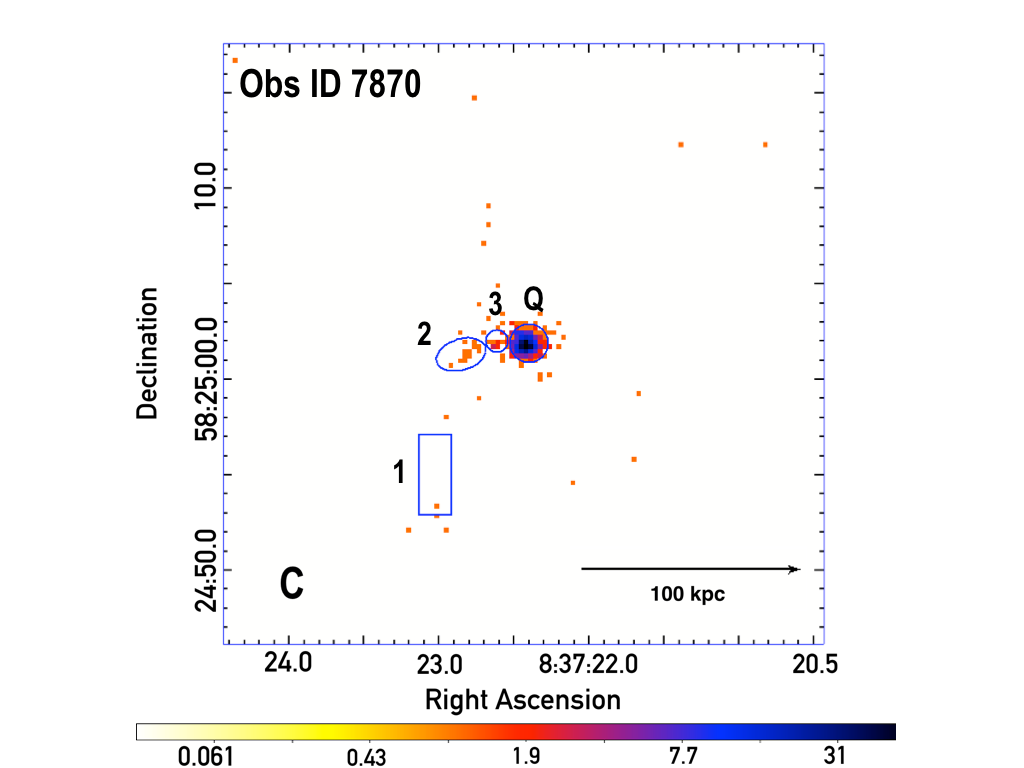}\\
\end{tabular}
\quad
\begin{tabular}[b]{@{}p{0.3\textwidth}@{}}
\centering\small (c)
\centering\includegraphics[scale=0.205,trim= 120 0 170 25,clip]{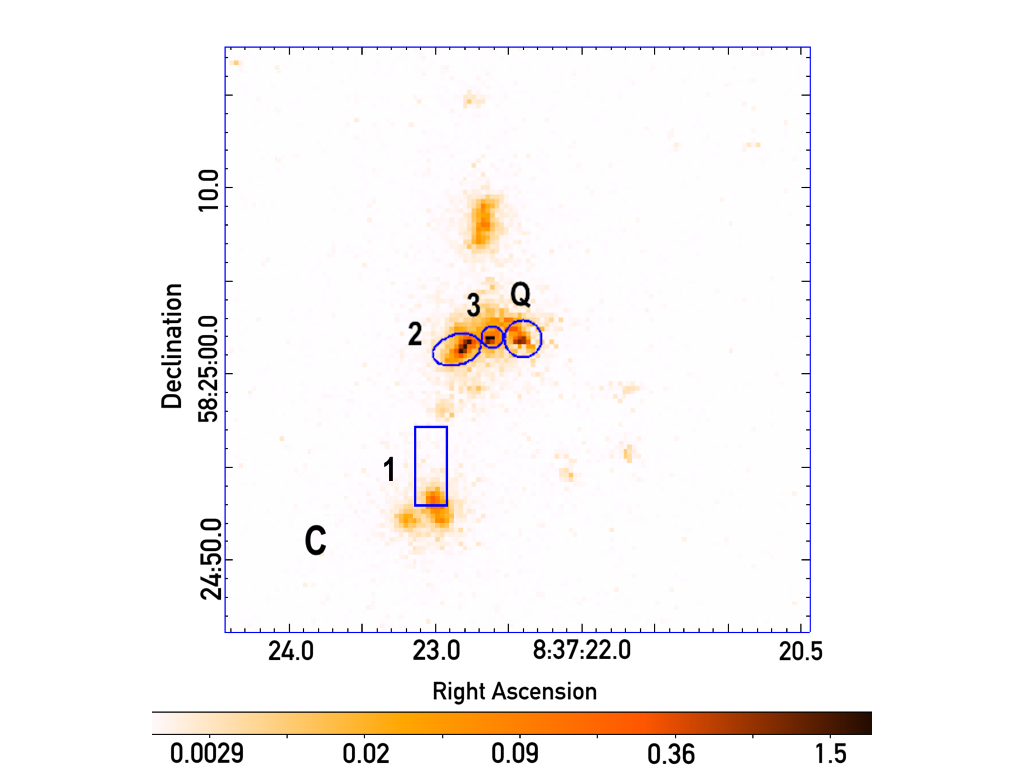}\\
\end{tabular}\\
\begin{tabular}[b]{@{}p{\textwidth}@{}}
\centering\small (d)\\
\centering\includegraphics[scale=0.5]{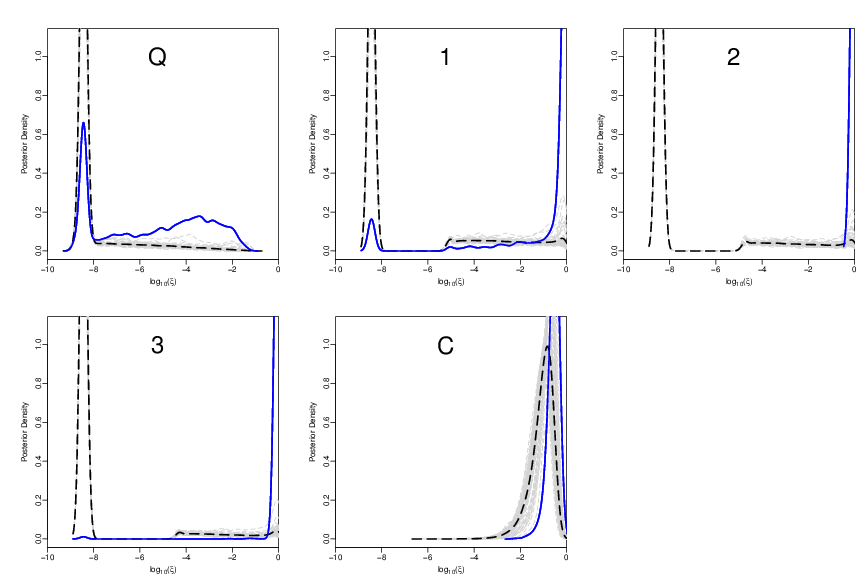}
\end{tabular}
\caption{As in Figure~\ref{fig:10307}, for quasar 0833$+$585 (ObsID 7870).}
\label{fig:7870}
\end{figure*}

\begin{figure*}[h]
\centering
\begin{tabular}[b]{@{}p{0.3\textwidth}@{}}
\centering\small (a)
\centering\includegraphics[scale=0.195,trim= 120 0 170 30,clip]{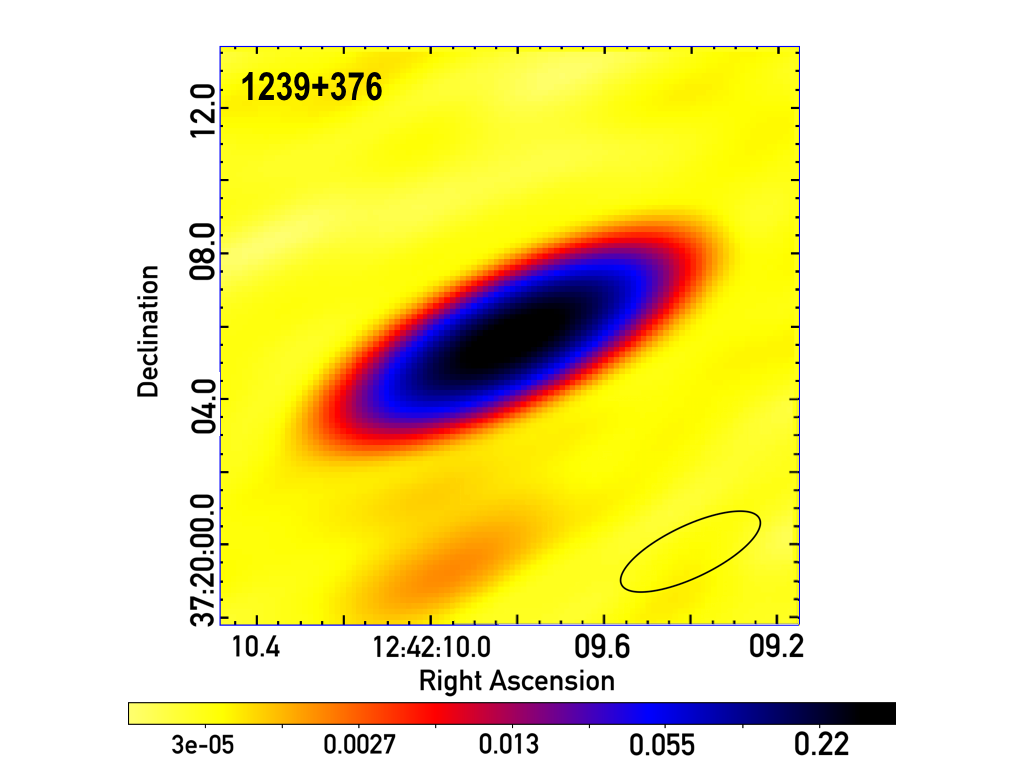}\\
\end{tabular}
\quad
\begin{tabular}[b]{@{}p{0.3\textwidth}@{}}
\centering\small (b)
\centering\includegraphics[scale=0.2,trim= 120 0 170 55,clip]{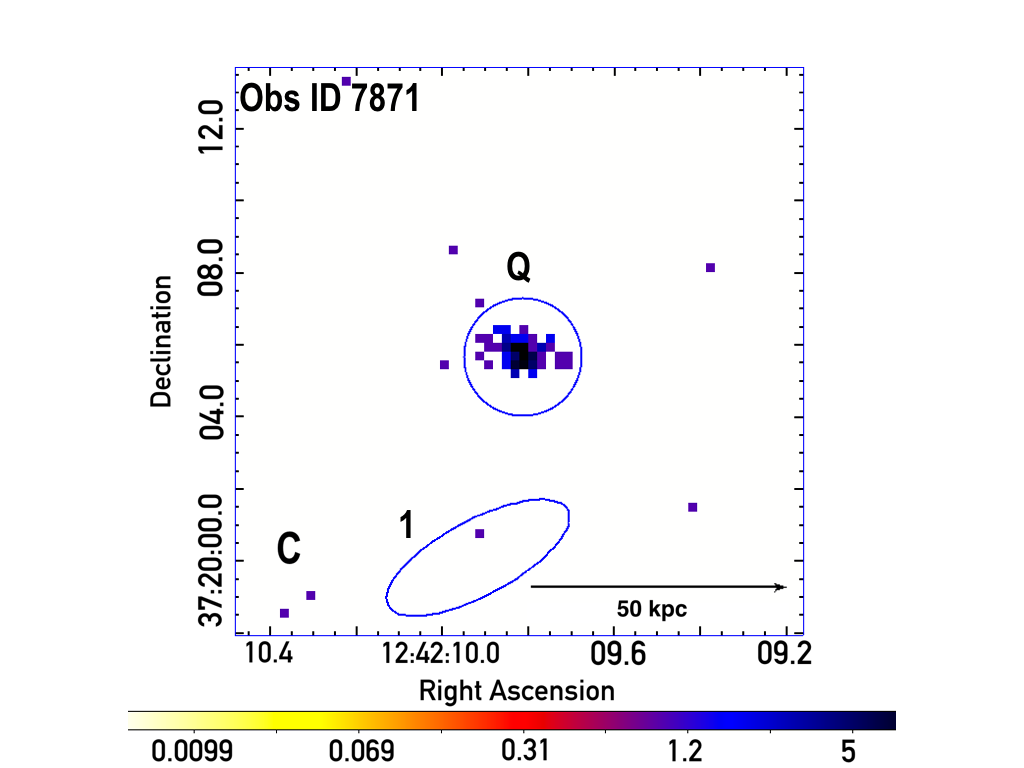}\\
\end{tabular}
\quad
\begin{tabular}[b]{@{}p{0.3\textwidth}@{}}
\centering\small (c)
\centering\includegraphics[scale=0.19,trim= 120 0 170 20,clip]{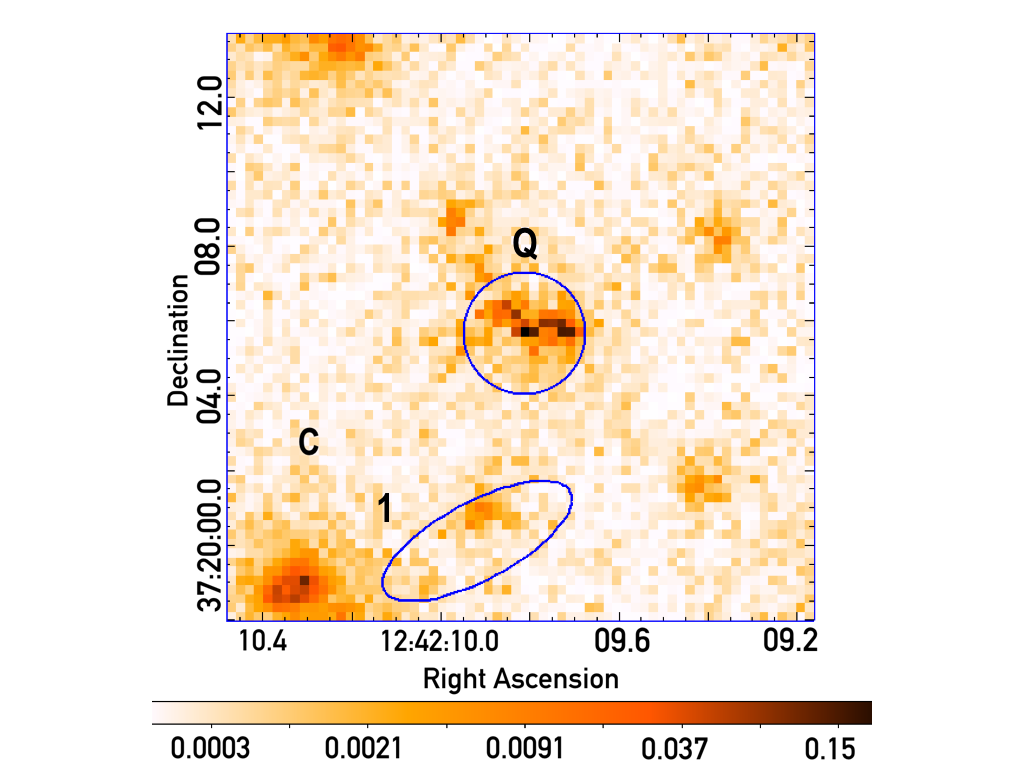}\\
\end{tabular}
\begin{tabular}[b]{@{}p{\textwidth}@{}}
\centering\small (d)\\
\centering\includegraphics[scale=0.5]{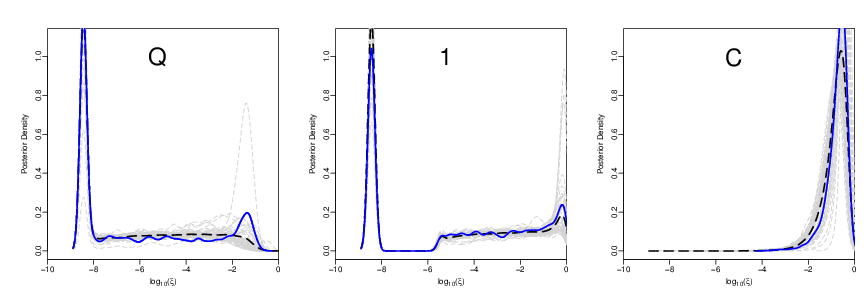}
\end{tabular}
\caption{As in Figure~\ref{fig:10307}, for quasar 1239$+$376 (ObsID 7871).}
\label{fig:7871}
\end{figure*}

\begin{figure*}[h]
\centering
\begin{tabular}[b]{@{}p{0.3\textwidth}@{}}
\centering\small (a)
\centering\includegraphics[scale=0.205,trim= 120 0 170 38,clip]{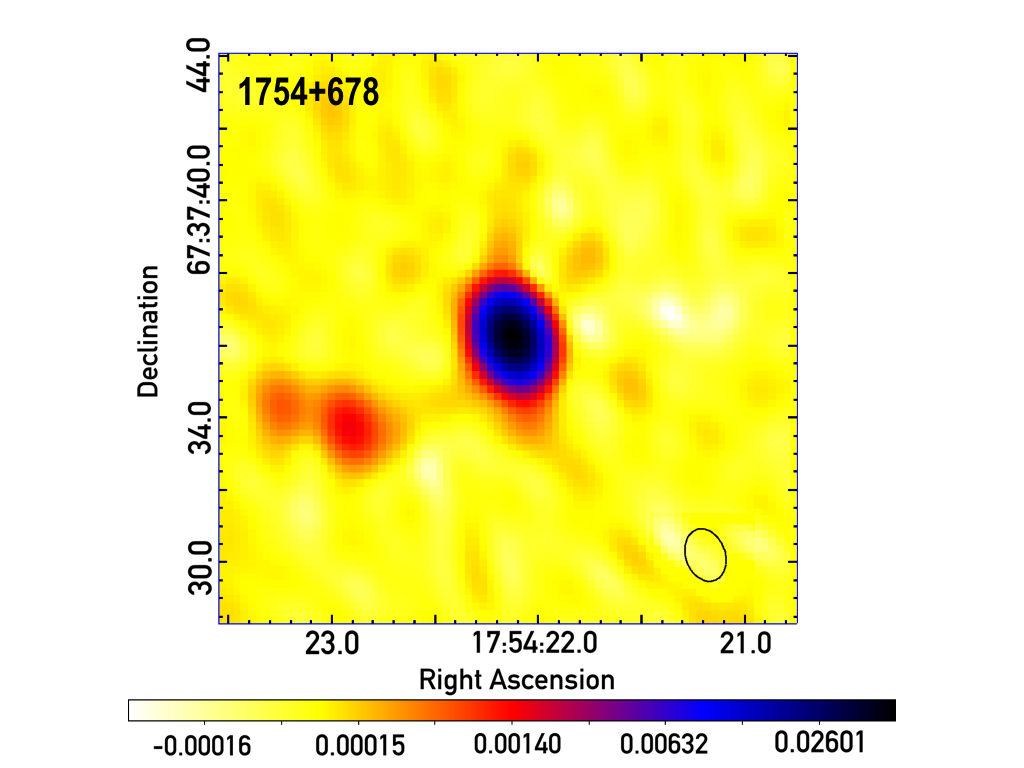}\\
\end{tabular}
\quad
\begin{tabular}[b]{@{}p{0.3\textwidth}@{}}
\centering\small (b)
\centering\includegraphics[scale=0.202,trim= 120 0 150 50,clip]{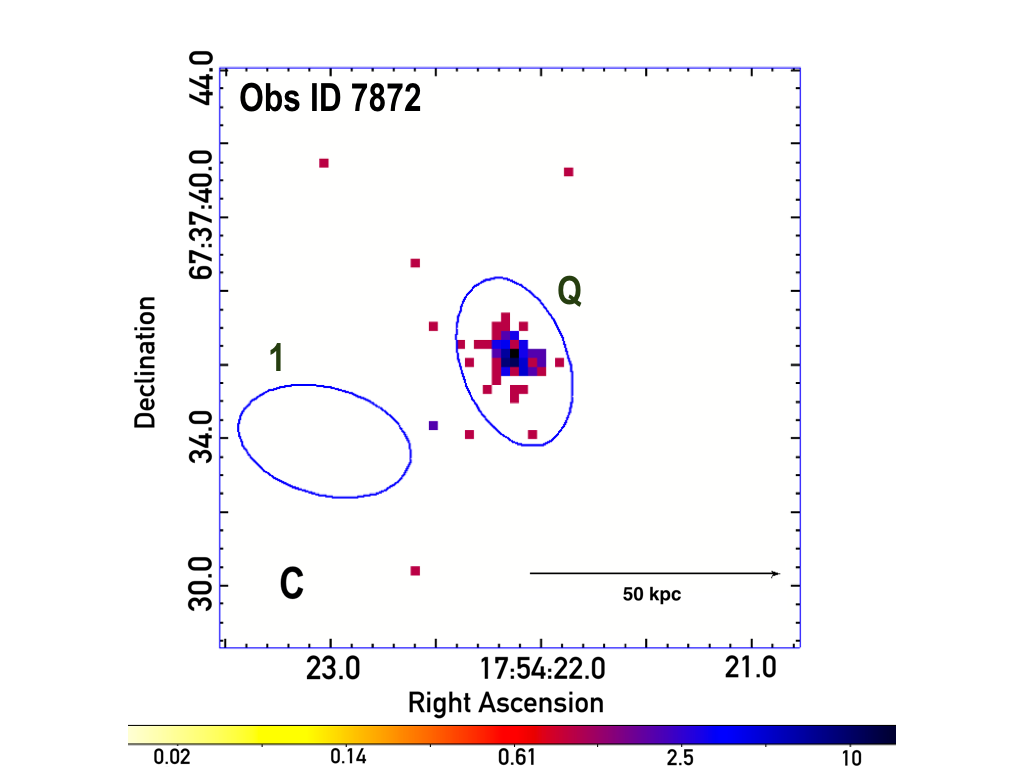}\\
\end{tabular}
\quad
\begin{tabular}[b]{@{}p{0.3\textwidth}@{}}
\centering\small (c)
\centering\includegraphics[scale=0.195,trim= 120 0 150 15,clip]{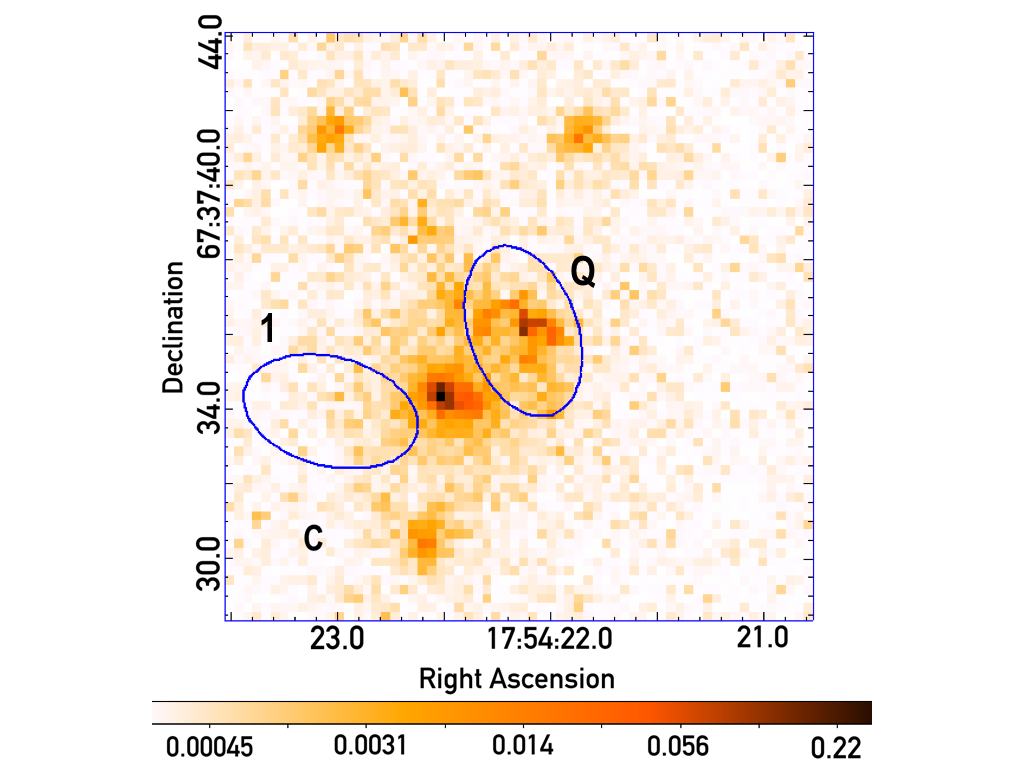}\\
\end{tabular}\\
\begin{tabular}[b]{@{}p{\textwidth}@{}}
\centering\small (d)\\
\centering\includegraphics[scale=0.5]{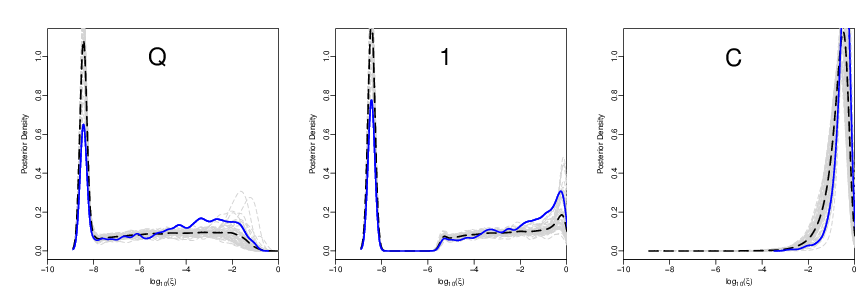}
\end{tabular}
\caption{As in Figure~\ref{fig:10307}, for quasar 1754$+$678 (ObsID 7872).}
\label{fig:7872}
\end{figure*}

\begin{figure*}[h]
\centering
\begin{tabular}[b]{@{}p{0.3\textwidth}@{}}
\centering\small (a)
\centering\includegraphics[scale=0.2,trim= 120 0 170 35,clip]{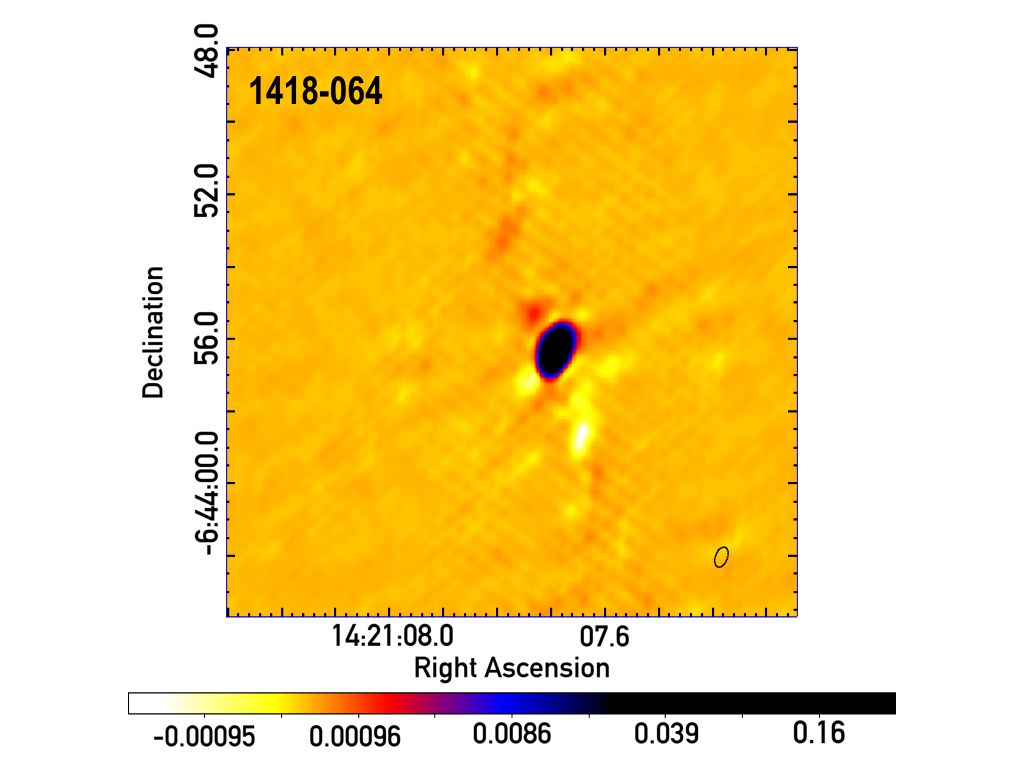}\\
\end{tabular}
\quad
\begin{tabular}[b]{@{}p{0.3\textwidth}@{}}
\centering\small (b)
\centering\includegraphics[scale=0.2,trim= 120 0 170 60,clip]{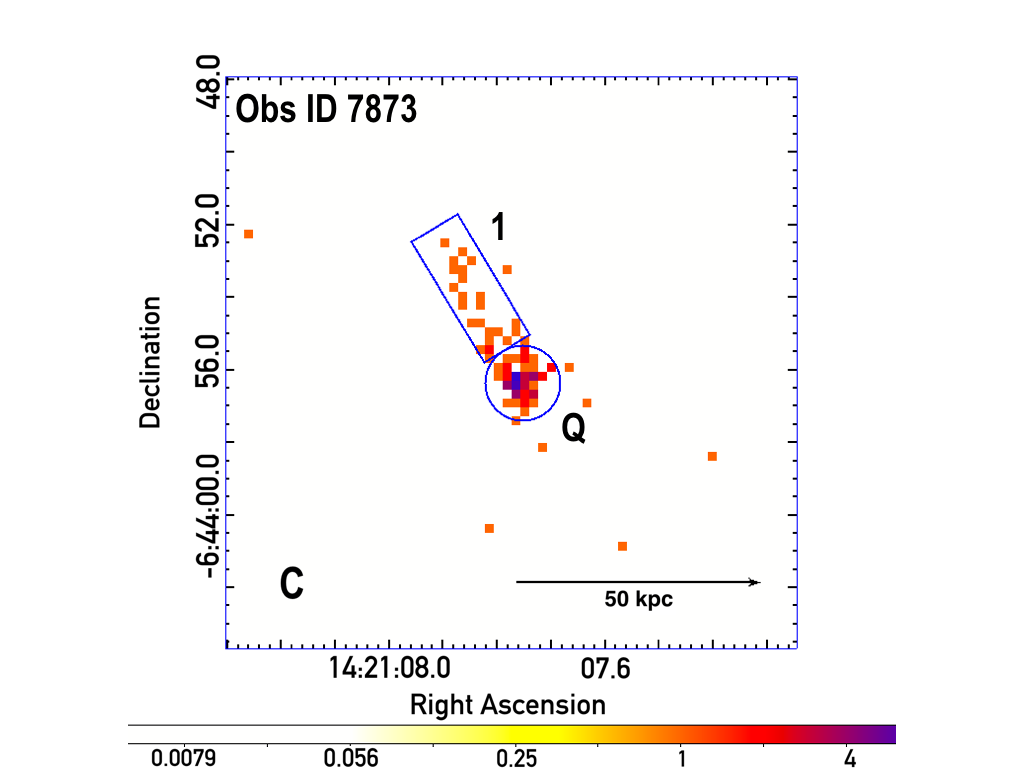}\\
\end{tabular}
\quad
\begin{tabular}[b]{@{}p{0.3\textwidth}@{}}
\centering\small(c)
\centering\includegraphics[scale=0.195,trim= 120 0 170 20,clip]{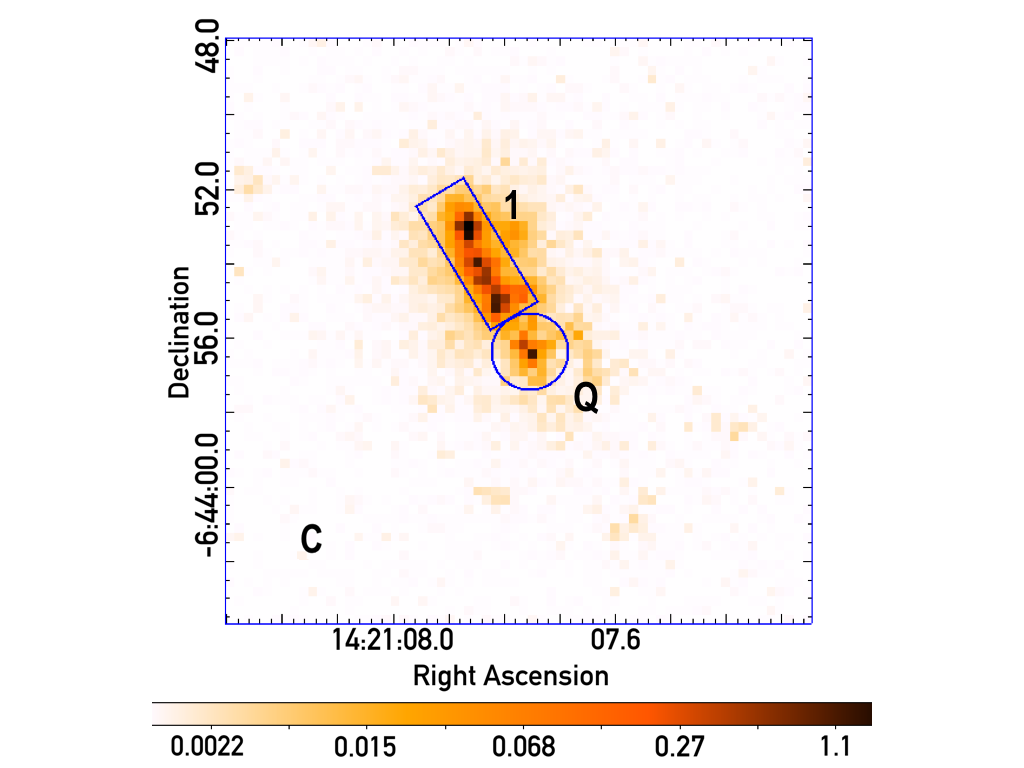}\\
\end{tabular}\\
\begin{tabular}[b]{@{}p{\textwidth}@{}}
\centering\small (d)\\
\centering\includegraphics[scale=0.5]{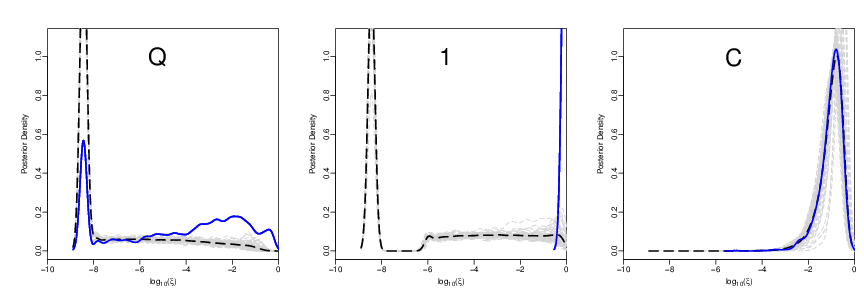}
\end{tabular}
\caption{As in Figure~\ref{fig:10307}, for quasar 1418$-$064 (ObsID 7873).}
\label{fig:7873}
\end{figure*}

\begin{figure*}[h]
\centering
\begin{tabular}[b]{@{}p{0.3\textwidth}@{}}
\centering\small (a)
\centering\includegraphics[scale=0.2,trim= 120 0 170 35,clip]{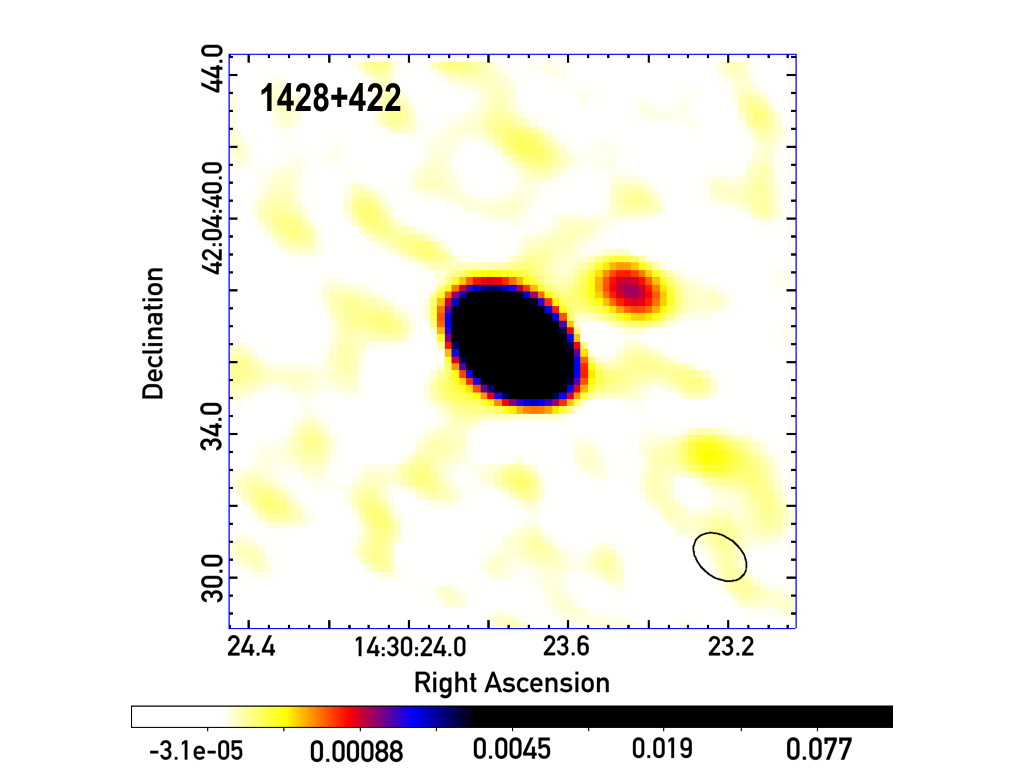}\\
\end{tabular}
\quad
\begin{tabular}[b]{@{}p{0.3\textwidth}@{}}
\centering\small (b)
\centering\includegraphics[scale=0.2,trim= 130 0 170 35,clip]{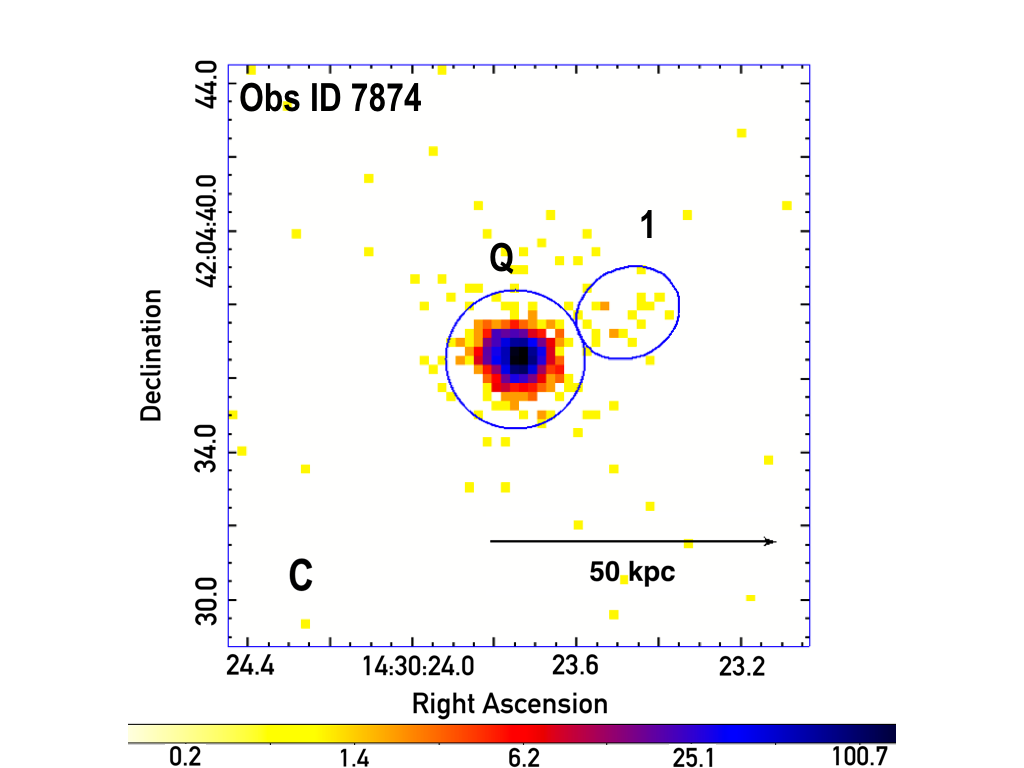}\\
\end{tabular}
\quad
\begin{tabular}[b]{@{}p{0.3\textwidth}@{}}
\centering\small (c)
\centering\includegraphics[scale=0.195,trim= 120 0 170 10,clip]{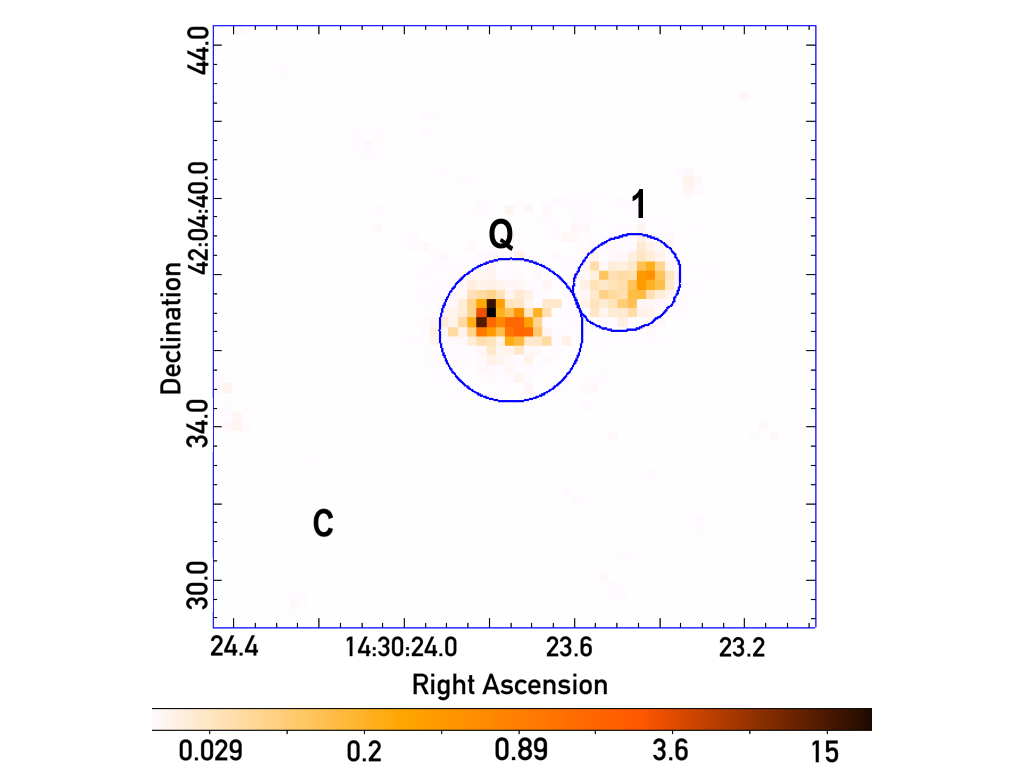}\\
\end{tabular}\\
\begin{tabular}[b]{@{}p{\textwidth}@{}}
\centering\small (d)\\
\centering\includegraphics[scale=0.5]{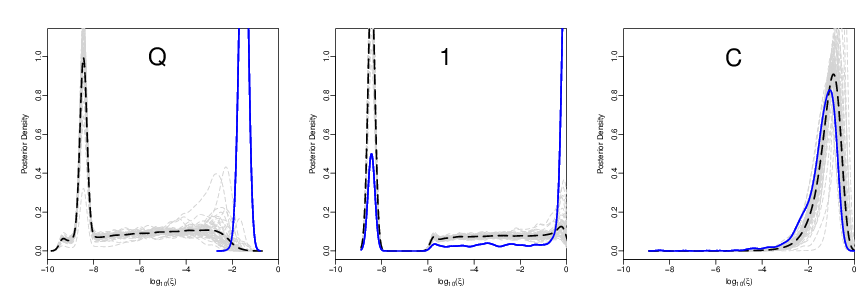}
\end{tabular}
\caption{As in Figure~\ref{fig:10307}, for quasar 1428$+$422 (ObsID 7874).}
\label{fig:7874}
\end{figure*}

\begin{figure*}[h]
\centering
\begin{tabular}[b]{@{}p{0.3\textwidth}@{}}
\centering\small (a)
\centering\includegraphics[scale=0.2,trim= 120 0 170 30,clip]{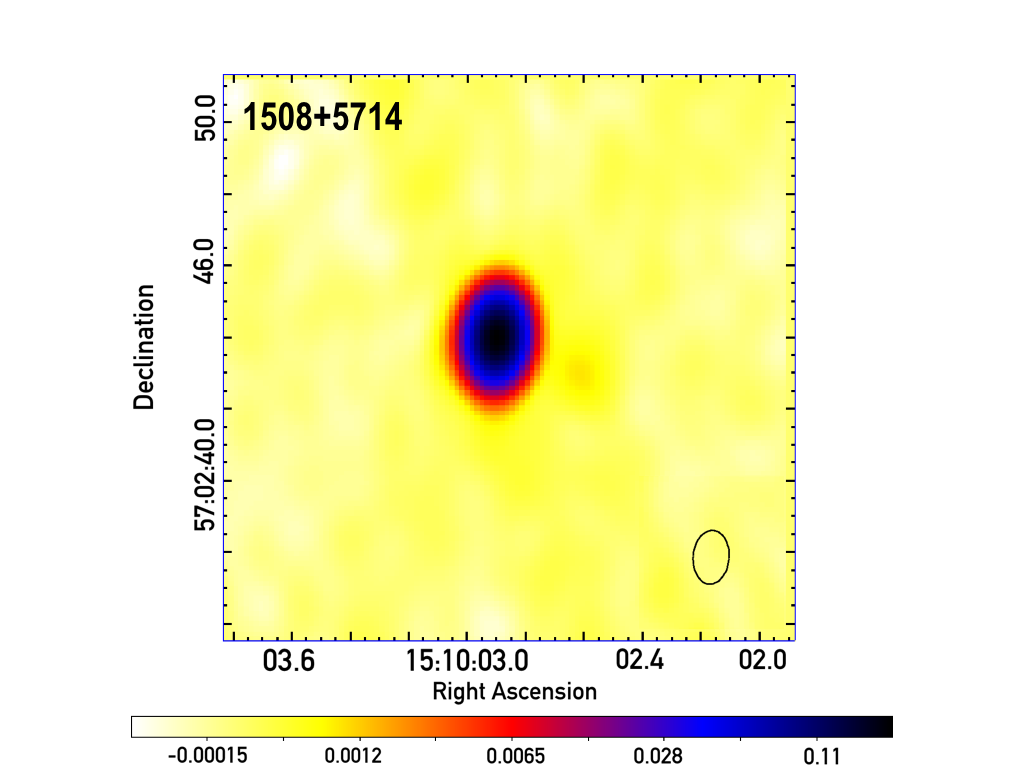}\\
\end{tabular}
\quad
\begin{tabular}[b]{@{}p{0.3\textwidth}@{}}
\centering\small (b)
\centering\includegraphics[scale=0.2,trim= 120 0 170 20,clip]{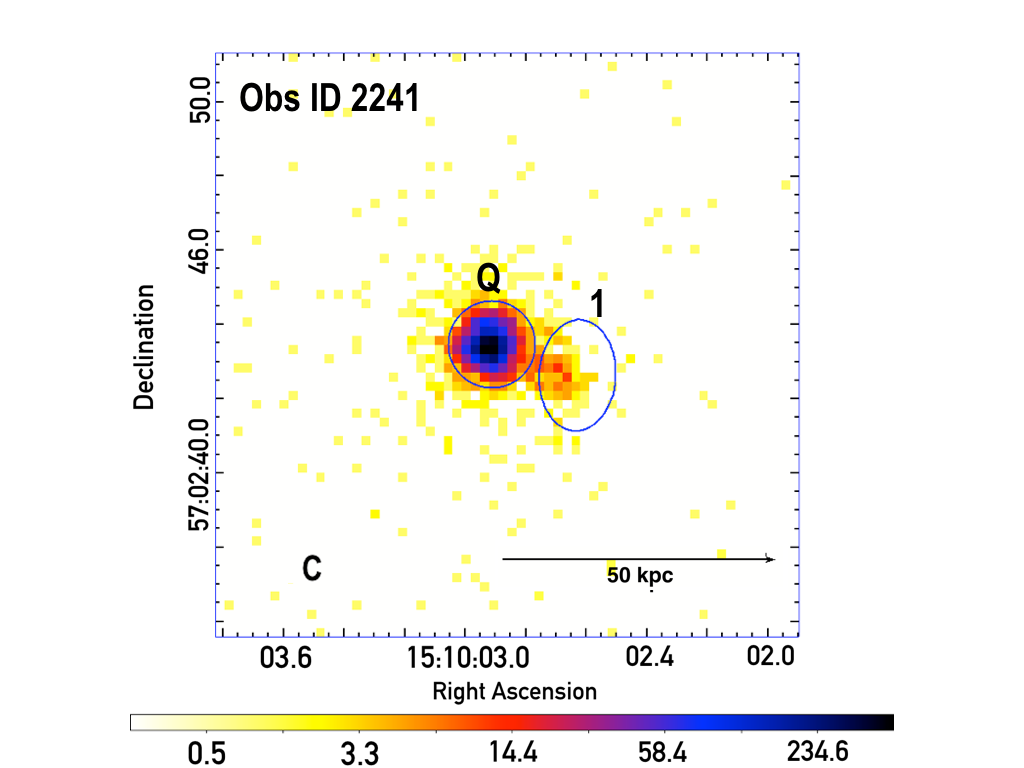}\\
\end{tabular}
\quad
\begin{tabular}[b]{@{}p{0.3\textwidth}@{}}
\centering\small (c)
\centering\includegraphics[scale=0.2,trim= 120 0 170 20,clip]{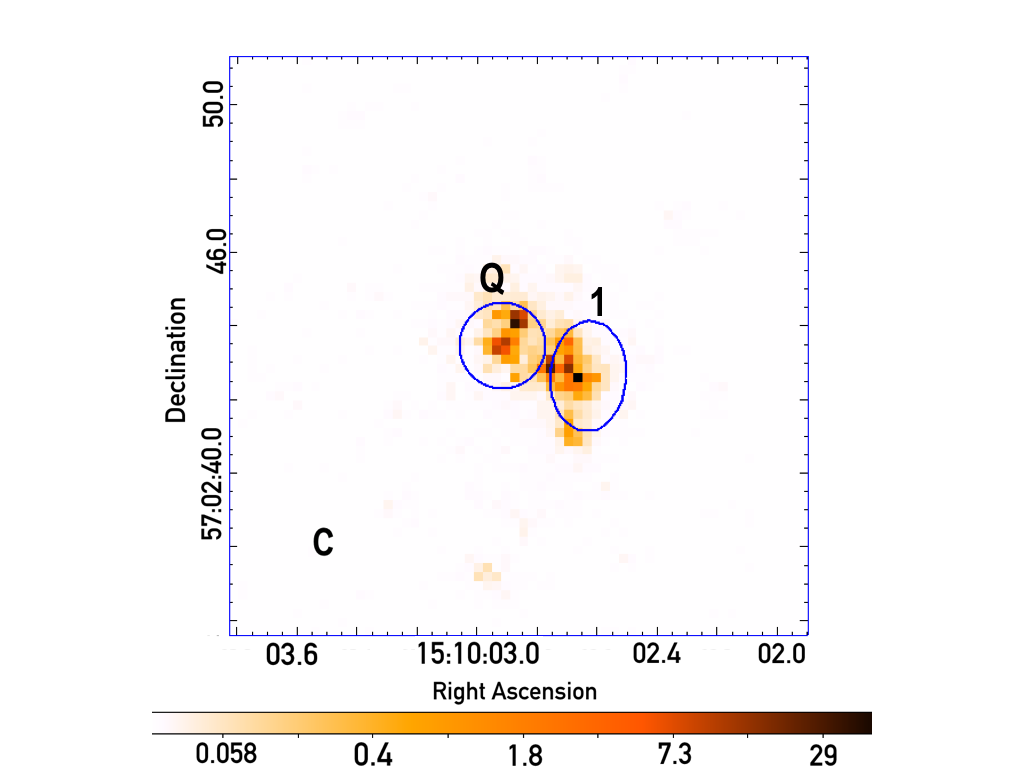}\\
\end{tabular}\\
\begin{tabular}[b]{@{}p{\textwidth}@{}}
\centering\small (d)\\
\centering\includegraphics[scale=0.5]{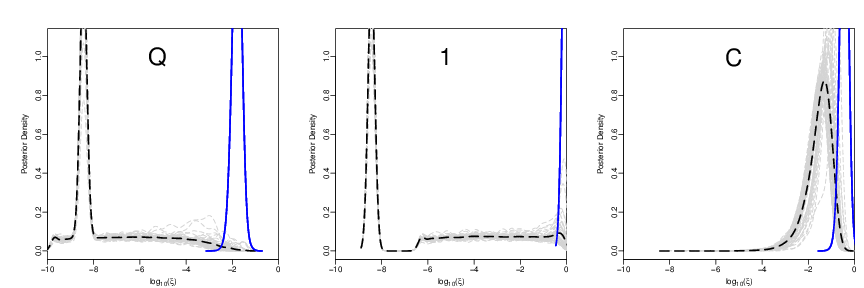}
\end{tabular}
\caption{As in Figure~\ref{fig:10307}, for quasar 1508$+$5714 (ObsID 2241).}
\label{fig:2241}
\end{figure*}
\end{document}